\documentclass[twocolumn]{aa}
\usepackage{graphicx}
\usepackage{longtable,lscape}
\usepackage{txfonts}
\begin{document}

\title{Spectroscopic and Photometric evidence of two stellar
  populations in the Galactic Globular Cluster NGC\ 6121 (M4)
\thanks{Based  on data collected at the European Southern
    Observatory with the VLT-UT2, Paranal, Chile.} }

%\subtitle{Chemical abundances of RGB stars from UVES spectra. }

\author{A.\ F.\ Marino
        \inst{1},
	S.\ Villanova
        \inst{2},
	G.\ Piotto
        \inst{1},
        A.\ P.\ Milone
        \inst{1},
        Y.\ Momany
        \inst{3},
        L.\ R.\ Bedin
        \inst{4}
        \and
        A.\ M.\ Medling
        \inst{5}
          }

%\offprints{XXX.\ XXX.}

\institute{Dipartimento  di Astronomia,  Universit\`a  di  Padova,
	   vicolo dell'Osservatorio 2, Padova, I-35122, Italy, EU\\
           \email{anna.marino-giampaolo.piotto-antonino.milone@unipd.it}
           \and
           Departamento de Astronomia, Universidad de Concepcion, Casilla
	   160-C, Concepcion, Chile\\
           \email{svillanova@astro-udec.cl}
           \and
           Osservatorio Astronomico di Padova, Vicolo dell'Osservatorio 5, 
           35122 Padova, Italy\\
           \email{yazan.almomany@oapd.inaf.it}
           \and
           Space Telescope Science Institute, Baltimore, MD 21218, USA\\
	   \email{bedin@stsci.edu}
           \and
           Department of Astronomy and Astrophysics, University of
           California, Santa Cruz, CA, 95064\\
           \email{amedling@ucolick.org}
           }

\date{Received Xxxxx xx, xxxx; accepted Xxxx xx, xxxx}

%__________________________________________________________________
%

% \abstract{}{}{}{}{} 
% 5 {} token are mandatory
 
  \abstract
   {}
   {We present abundance analysis based on high resolution spectra
    of 105 isolated red giant branch (RGB) stars in the Galactic 
    Globular Cluster NGC~6121 (M4). Our aim is to study its star population
    in the context of the multi-population phenomenon recently discovered
    to affect some Globular Clusters.}
   {The data have been collected with FLAMES+UVES, the multi-fiber 
    high resolution facility at the ESO/VLT@UT2 telescope. Analysis was
    performed under LTE approximation for the following elements: O, Na,
    Mg, Al, Si, Ca, Ti, Cr, Fe, Ni, Ba, and NLTE corrections were applied 
    to those (Na, Mg) strongly affected by departure from LTE.
    Spectroscopic data were coupled with high-precision wide-field UBVI$_{\rm C}$ 
    photometry from WFI@2.2m telescope  and infrared JHK photometry from 2MASS.}
   {We derived an average $\rm {[Fe/H]}=-1.07\pm0.01$ (internal error), and
    an $\alpha$ enhancement of $\rm {[\alpha/Fe]}=+0.39\pm0.05$ dex (internal error).
    We confirm the presence of an extended Na-O anticorrelation, and find
    two distinct groups of stars with significantly different Na and O content.
    We find no evidence of a Mg-Al anticorrelation.
    By coupling our results with previous studies on the CN band strength, 
    we find that the CN strong stars have higher Na and Al content and are 
    more O depleted than the CN weak ones.
    The two groups of Na-rich, CN-strong and Na-poor, CN-weak stars populate two 
    different regions along the RGB.
    The Na-rich group defines a narrow sequence on the red side of the 
    RGB, while the Na-poor sample populate a bluer, more spread portion of the RGB.
    In the $\rm {U}$ vs.\ $\rm {U-B}$ color magnitude diagram the RGB
    spread is present from the base of the RGB to the RGB-tip.
    Apparently, both spectroscopic and photometric results imply the presence
    of two stellar populations in M4. We briefly discuss the possible
    origin of these populations.}
   {}

   \keywords{Spectroscopy ---
          globular clusters: individual: NGC~6121}

\titlerunning{Spectroscopic and Photometric study of the Globular Cluster M4.}
\authorrunning{Marino et al.}

\maketitle
%
%________________________________________________________________

\section{Introduction}

Observational evidence for variations in the chemical composition of
light elements in Globular Cluster (GC) stars were known since Cohen (1978),
who noted a scatter in Na among stars in M3 and M13.
During the last few decades, high resolution spectroscopic
studies have definitely confirmed that a GC stellar population is not
chemically homogeneous. Even if GC stars are generally homogeneous in
their Fe-peak element content, they show large star-to-star abundance
variations in the light elements such as C, N, O, Na, Mg, Al, 
and others (see Gratton et al.\ 2004 for a review).

During the last twenty years, it has become clear that in red giant stars the
abundances of some of these elements follow a well defined pattern. In
particular, there are clear anticorrelations between the Na and O
content, and between Mg and Al. Variations in the molecular CH, CN
and NH band strengths, due to a spread in the abundances of carbon and
nitrogen have been observed, as well as anticorrelations between CH
and CN strengths, and, in some cases, a clear bimodality in the CN
content.

Despite the spectroscopic observational evidence collected in more
than thirty years, the pattern in the light elements is not yet well 
understood. Two scenarios have been proposed to explain
this observed chemical heterogeneity: the
evolutionary scenario and the primordial one, both apparently
supported by observations.

The observed decline of C content (e.g. in M13, as shown by Smith \&
Briley 2006) and the decreasing ratio $\phantom{}^{12}\rm
{C}/\phantom{}^{13}\rm {C}$ (observed in M4 and NGC~6528 by Shetrone 2003) 
along the RGB phase, support the evolutionary
scenario. According to this theory, the origin of the observed star-to-star
scatter in some elements is due to the mixing of CNO-cycle-processed
material transported, in a way not well understood yet, to the stellar
surface. In this way the observed anticorrelations would be present in
the evolved stages of the life of stars, after the RGB bump.

At odds with this scenario, in the last few years, spectroscopic
studies have revealed light element abundance variations in
unevolved main sequence stars and less-evolved RGB stars, fainter
than the RGB bump. The Na-O anticorrelation was found at the level of
the main sequence turn-off (TO) and sub giant branch (SGB) in M13
(Cohen \& Mel\'endez 2005), NGC~6397 and NGC~6752 (Carretta et al.\ 2005;
Gratton et al.\ 2001), NGC~6838 (Ram\'irez \& Cohen 2002) and 47~Tuc
(Carretta et al.\ 2004). In 47~Tuc, a bimodal distribution in the CN
strengths, similar to that found among RGB stars (Norris \& Freeman 1979), 
was found also in the main sequence (MS) by Cannon et al.\ (1998).  
Moreover, Grundahl et al.\ (2002) have shown that in
NGC~6752 the observed scatter in the Str$\rm {\ddot {o}}$mgren index
$c_{1}$ is due to the abundance variations in NH bands in stars both
brighter and fainter than the RGB bump. This result is consistent with a
primordial scenario since theory does not predict significant mixing
below the luminosity of the first dredge-up, observationally
corresponding to the magnitude of the RGB bump. These observations
suggest that the light element variations should be primordial,
i.e. they are derived from the chemical composition of the primordial site
where the GC stars have been generated, or alternatively, that a second
generation of stars has been formed from a medium enriched in some
elements (e.g., the self enrichment model by Ventura et al.\ 2001).
A primordial scenario in which such GCs have experienced multiple
episodes of star formation challenges the paradigm that GCs host a
single stellar population, i.e. that stars of a given cluster are
coeval and chemically homogeneous.

Very recently, a spectacular, and somehow unexpected confirmation
that, at least in some GCs, the origin of the chemical anomalies must
be primordial came from high precision photometry from Hubble Space
Telescope observations. The first object challenging the paradigm of
GCs hosting simple stellar populations was $\omega$~Cen. As shown by
Bedin et al.\ (2004), the MS of $\omega$~Cen is split into two
sequences. But the most exciting discovery came from the
spectroscopic investigation by Piotto et al.\ (2005), who found that
the bluest MS is more metal rich than the redder one. The only way to
account for the spectroscopic and photometric properties of the two
main sequences is to assume that the bluest sequence is also strongly
He enhanced, to an astonishingly high Y=0.38.  More recently,
Villanova et al.\ (2007) showed that the two main sequences split in at
least four sub giant branches (SGB) which must be connected in some way
to the multiplicity of RGBs identified by Lee et al.\ (1999) and
Pancino et al.\ (2000). The metal content of the different SGBs
measured by Villanova et al.\ (2007) also implies a large age
difference among $\omega$~Cen stellar populations, larger than
1 Gyr. The exact age dispersion depends on the assumed abundances (including
the He content) of the different stellar populations, and it is still
controversial (Sollima et al.\ 2007). Villanova et al.\ (2007) demonstrated
that there is also a third MS, running on the red side of the two main MSs,
and likely connected with the anomalous RGB-a of Pancino et al.\ (2000).  

Omega Centauri was well-known since the seventies because of its peculiar
metallicity distribution. It is the only GC showing iron-peak element
dispersion (Freeman \& Rodgers 1975 and, more recently, Norris et al.\ 1996,
Suntzeff \& Kraft 1996). In a sense, the findings by Pancino et
al.\ (2000), Bedin et al.\ (2004), Piotto et al.\ (2005), Villanova et
al.\ (2007) and Sollima et al.\ (2007) could simply be considered
additional evidence that $\omega$~Cen is so peculiar that it might
not be a GC. Perhaps, as suggested by many authors
(Freeman 1993, Hughes \& Wallerstein 2000), it might simply be the nucleus 
of a much larger system, likely disrupted by the tidal field of our Galaxy.\\
In this sense the most recent discovery by Piotto et al.\ (2007) that the MS
of NGC~2808 is split into three, distinct sequences came as a
sort of surprise, shaking at its foundation our understanding of GC
stellar populations. NGC~2808 has been always considered a GC, with
many peculiar properties regarding its metal content and its
color-magnitude diagram, but a genuine, massive GC. Still, it hosts multiple
stellar populations.  Moreover, also in this case, in view of
the negligible dispersion in iron-peak
elements in NGC~2808, the only way so far available to reproduce the
three MSs is to assume that there are three populations, characterized
by three different helium contents, up to an (again) astonishingly high
Y=0.40. Interestingly enough, D'Antona et al.\ (2005) already made the
hypothesis of three groups of stars, with three different helium
abundances in order to explain the peculiar, multi-modal Horizontal Branch
(HB) of NGC~2808 (Sosin et al.\ 1997, Bedin et al.\ 2000).
Piotto et al.\ (2007) simply found them in the form of a MS split.
Piotto et al.\ (2007) also noticed that the different stellar
populations in NGC~2808 are consistent with the spectroscopic
observations by Carretta et al.\ (2006), who identified three groups of
stars with different Oxygen abundances. The fraction of stars in the
three abundance groups is in rough agreement with the fraction of stars
in the three MSs.

NGC~2808 and $\omega$~Cen are, at the moment, the most extreme examples
of a rather complex observational scenario. In fact, evidence of
multiple populations has been found in other GCs, like NGC~1851
(Milone et al.\ 2008), NGC~6388 (Siegel et al.\ 2007, Piotto et al.\
2008), and M54 (Piotto et al., in preparation) in the form 
of a split in the SGB. In NGC~1851 the presence of a group of RGB
stars with enhanced Sr and Ba and strong CN bands, among the majority
of CN-normal RGB stars (Yong \& Grundahl 2008), and the presence of a bimodal
HB, agrees with the hypothesis of two stellar generations inferred by
the observed SGB split.

The extremely peculiar HB of NGC~6388 (Rich et al.\ 1997), with the
presence of extremely hot HB stars (Busso et al.\ 2007) was already
interpreted in terms of multiple, helium enhanced, population by
Sweigart \& Catelan (1998), and in the more detailed analysis by
Caloi \& D'Antona (2007).

As for M54, it has been recognized as the nucleus of the Sagittarius
dwarf galaxy (Da Costa \& Armandroff 1995; Bassino \& Muzzio 1995;
Layden \& Sarajedini 2000), and it could simply represent what
$\omega$~Cen was long time ago. 

It is worth noting that the clusters showing multiplicities in their CMDs are among the most
massive GCs of our Galaxy ($\rm {M > 10^6 M_\odot}$) and all of them have
peculiar HBs, as well as peculiar abundances, including Na-O
anticorrelations.
However, we must also note that much less massive GCs, with no
evidence of multiple populations identified so far, show very large
star-to-star abundance variations. It is noteworthy to recognize here
that the Na-O anticorrelation has been found in about 20 GCs (see
Carretta et al.\ 2006 for the most updated list).

In summary, it is clear that GCs are not as simple systems as thought
in the past. Up to now, we lack a complete explanation for the mechanisms
necessary to understand the observational scenario. A
systematic study of the chemical abundances of many stars in GCs is
needed in order to better understand the star formation history of
these objects.

In this work we present a study on the chemical abundances of the GC
NGC~6121 (M4) from high resolution spectra of its RGB
stars.

As far as we know, M4 shows no evidence of multiple stellar
populations in its CMD, and its mass ($\rm {log {\frac
{M}{M\odot}}=4.8}$, Mandushev et al.\ 1991) is much smaller than the
mass of the clusters with the photometric peculiarities discussed above.

Chemical abundances from high resolution spectra of M4 RGB stars
have been already measured by different groups of investigators:
Gratton, Quarta \& Ortolani 1986 (hereafter GQO86), Brown \&
Wallerstein 1992 (BW92), Drake et al.\ 1992 (D92), 
Ivans et al.\ 1999 (I99), and Smith et al.\ (2005).
These authors have found a range of [Fe/H] between $-$1.3 dex
(BW92) and $-$1.05 dex (D92). For further details, see I99 who have summarized
the chemical abundances found in previous studies.
A study by Norris 1981 (hereafter N81) of 45 RGB stars showed a CN
bimodality in M4, e. g., stars of very similar magnitudes and colors
have a bimodal distribution of CN band strengths. 
By analyzing 4 RGB stars, two selected from the CN-weak group and two from the
CN-strong one, D92 (and then Drake et al.\ 1994)
found differences in the Na and Al content. I99 found that O is
anticorrelated with N, whereas Na and Al abundances are larger in
CN-strong stars. Looking at the evolutionary states of these stars, both I99 and Smith
\& Briley 2005 (SB05) did not find any strong correlation between CN band
strength and the position on the CMD. More recently, Smith et al.\ (2005),
studying the fluorine abundance in seven RGB stars in M4, found a large
variation in $\phantom{}^{19}$F, which is anticorrelated with the Na and Al
abundances. In these previous studies, both the evolutionary scenario and
a primordial one have been taken into account in order to explain the
light element variations and the CN bimodality in M4.

In this work, we analyze high resolution spectra in order to study
chemical abundances for a large sample of M4 RGB stars and compare our results
with those of previous studies.
In Section 2 we provide an overview of the observations and target sample.
The membership criterion used to separate the probable cluster stars is
described in Section 3, and the procedure to derive the chemical
abundances is in Section 4. We present our results in Section 5, and
discuss them in Section 6 and Section 7. Section 8 summarizes the results of this work.

\begin{figure}[hpbt]
\centering
\includegraphics*[width=9.5cm]{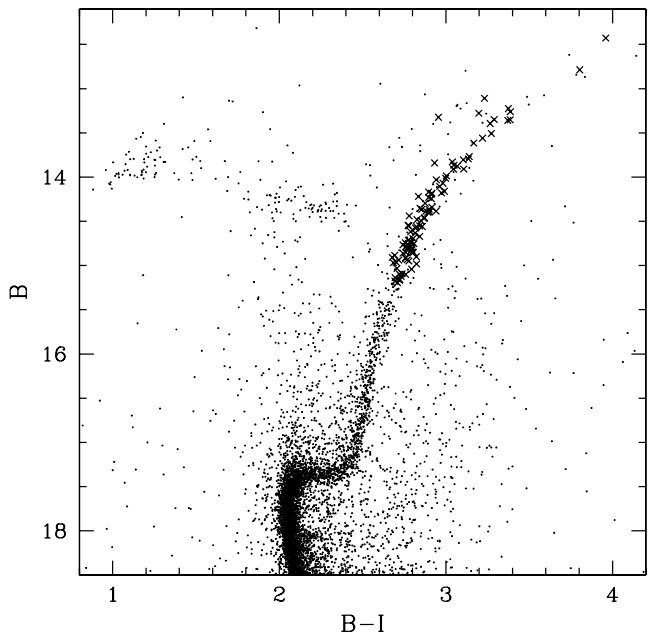}
\caption{Distribution of the UVES target
         stars on the $\rm {B}$ vs.\ $\rm {(B-I)}$ CMD corrected for
         differential reddening.} 
\label{Fig1}
\end{figure}

\section{Observations and Data Reduction}

Our dataset consists of UVES spectra collected in
July-September 2006, within a project devoted to the
detection of spectroscopic binaries in the GC M4, and to the
measurement of the geometric
distance of this cluster (Programs 072.D-0742 and 077.D-0182).  Data are
based on single 1200-1800 sec exposures obtained with FLAMES/VLT@UT2
(Pasquini et al.\ 2002) under photometric conditions and a typical seeing of
0.8-1.2 arcsec. 
The 8 fibers feeding the UVES spectrograph were centered on 115 isolated stars
(no neighbours within a radius of 1.2 arcsec brighter than $\rm {V+2.5}$, 
where $\rm {V}$ is the magnitude of the target star)
from $\sim1$ mag below the HB to the tip of the RGB of M4, in the magnitude
range $10.5<{\rm V}<14.0$.

The UVES spectrograph was used in
the RED 580 setting. The spectra have a spectral coverage of
$\sim$2000 \AA \ with the central wavelength at 5800 \AA. The typical
signal to noise ratio is ${\rm S/N\sim 100-120}$.

Data were reduced using UVES pipelines (Ballester et al.\ 2000),
including bias subtraction, flat-field correction, wavelength
calibration, sky subtraction and spectral rectification. Stars
were carefully selected from high quality photometric $\rm {UBVI_{C}}$
observations by Momany et al.\ (2003)
obtained with the Wide Field Imager (WFI) camera at the ESO/MPI 2.2m
telescope (total field of view $34\times33~{\rm arcmin}^{2}$) coupled
with infrared JHK 2MASS photometry (Skrutskie et al.\ 2006). Given
the spatially variable interstellar reddening across the cluster
(Cudworth \& Rees 1990; Drake et al.\ 1994; Kemp et al.\ 1993; Lyons
et al.\ 1995; and I99) we have corrected our CMDs for this effect (as
done in Sarajedini et al.\ 2007).
Fig.~\ref{Fig1} shows the position of the target stars on the 
corrected B vs (B-I) CMD. 

\section{Radial velocities and membership}

In the present work, radial velocities were used as the membership criterion
since the cluster stars all have similar motion with respect to the observer.
The radial velocities (${\rm v_{r}}$) of the
stars were measured using the IRAF FXCOR task, which cross-correlates
the object spectrum with a template.  As a template, we used a
synthetic spectrum obtained through the spectral synthesis code
SPECTRUM (see {\sf http://www.phys.appstate.edu/spectrum/spectrum.html} for
more details), using a Kurucz model atmosphere with roughly the mean
atmospheric parameters of our stars $\rm {T_{eff}=4500}$ K, $\rm {log(g)=2.0}$,
$\rm {v_{t}=1.4}$ km/s, $\rm {[Fe/H]=-1.10}$. At the end, each radial velocity was
corrected to the heliocentric system. We calculated a first approximation
mean velocity and the r.m.s ($\sigma$) of the velocity distribution.
Stars showing $\rm {v_{r}}$ out of more than
$3\sigma$ from the mean value were considered probable field objects and
rejected, leaving us with 105 UVES spectra of probable members.
After this procedure, we obtained a new mean radial velocity of
$70.6\pm1$ km/s from all the selected spectra,
which agrees well with the values in the literature (Harris 1996, 
Peterson et al.\ 1995).

\section{Abundance analysis}

The chemical abundances for all elements, with the exception of Oxygen,
were obtained from the equivalent widths (EWs) of the spectral lines.  
An accurate measurement of EWs first requires a good determination
of the continuum level. Our relatively metal-poor stars, combined with our
high S/N spectra, allowed us to proceed in the following way. First, 
for each line, we selected a region of 20 \AA \ centered on the line itself
(this value is a good compromise between having enough points, i. e. a good statistic, and 
avoiding a too large region where the spectrum can be not flat).
Then we built the histogram of the distribution of the flux where the peak is a
rough estimation of the continuum. We refined this determination by fitting a
parabolic curve to the peak and using the vertex as our continuum estimation. 
Finally, the continuum determination was revised by eye and corrected by hand
if a clear discrepancy with the spectrum was found.
Then, using the continuum value previously obtained, we fit a gaussian curve
to each spectral line and obtained the EW from integration.
We rejected lines if affected by bad continuum determination, by non-gaussian
shape, if their central wavelength did not agree with that expected from 
our linelist, or if the lines were too broad or too narrow with respect to the
mean FWHM.
We verified that the gaussian shape was a good approximation for our spectral
lines, so no lorenzian correction was applied.
The typical error for these measurements is 3.6 m\AA,
as obtained from the comparison of the EWs measured on stars 
having similar atmospheric parameters (see Sect. 5.1 for further details).

\subsection{Initial atmospheric parameters}

Initial estimates of the atmospheric parameters were derived from WFI
$\rm {BVI}$ photometry. First-guess effective temperatures ($\rm
{T_{eff}}$) for each star were derived from the $\rm {T_{eff}}$-color relations
(Alonso et al.\ 1999).  
$\rm {(B-V)}$ and $\rm {(V-I)}$ colors were de-reddened using a reddening
of ${\rm E(B-V)=0.36}$ (Harris 1996). Surface gravities
($\rm {log(g)}$) were obtained from the canonical equation:

\begin{center}
${\rm log(g/g_{\odot}) = log(M/M_{\odot}) + 4\cdot
  log(T_{eff}/T_{\odot}) - log(L/L_{\odot}) }$
\end{center}

\noindent
where the mass ${\rm M/M_{\odot}}$ was derived from the spectral
type (derived from T$_{\rm eff}$) and the luminosity class of stars (in
this case we have a III luminosity class) through the grid of
Straizys \& Kuriliene (1981). 
The luminosity ${\rm L/L_{\odot}}$ was obtained from the absolute
magnitude ${\rm M_V}$, assuming an apparent distance modulus 
of $\rm {(m-M)_{V}=12.83}$ (Harris 1996). 
The bolometric correction ($\rm {BC}$) was derived by adopting the relation
$\rm {BC-T_{eff}}$ from Alonso et al.\ (1999). \\
Finally, microturbolence velocity ($\rm {v_{t}}$) was obtained from the
relation (Gratton et al. 1996):

\begin{center}
${\rm v_{t}\ (km/s) = 2.22-0.322\cdot log(g)}$
\end{center}

\noindent
These atmospheric parameters are only initial guesses and are
adjusted as explained in the following Section.

\subsection{Chemical abundances}

The Local Thermodynamic Equilibrium (LTE) program MOOG (freely
distributed by C. Sneden, University of Texas at Austin) was used to
determine the metal abundances.

The linelists for the chemical analysis were obtained from the VALD
database (Kupka et al.\ 1999) and calibrated using the Solar-inverse technique.
For this purpose we used the high resolution, high S/N Solar spectrum
obtained at NOAO ($National~Optical~Astronomy~Observatory$, Kurucz et
al.\ 1984). 
We used the model atmosphere interpolated from the Kurucz (1992) grid
using the canonical atmospheric parameters for the Sun: $\rm {T_{eff}=5777}$ K,
$\rm {log(g)}=4.44$, $\rm {v_{t}=0.80}$ km/s and $\rm {[Fe/H]=0.00}$.

The EWs for the reference Solar spectrum were obtained in the same
way as the observed spectra, with the exception of the strongest lines, where
a Voigt profile integration was used.
In the calibration procedure, we adjusted the value of the line
strength $\rm {log(gf)}$ of each
spectral line in order to report the abundances obtained from all the
lines of the same element to the mean value.
The chemical abundances obtained for the Sun
are reported in Tab.~\ref{Tab1}.
The derived Na and Mg abundances were corrected for the effects of
departures from the LTE assumption, using the prescriptions by Gratton
et al.\ (1999).

\begin{table} [!hbp]
\caption{Measured Solar abundances ($\rm
  {log\epsilon(X)=log(N_{X}/N_{H})+12)}$.} 
\label{Tab1}
\centering
\begin{tabular}{lcr}
\hline
\hline
Element & UVES         &lines\\
\hline
$\rm {OI}$     & 8.83         & 1  \\
$\rm {NaI_{NLTE}}$ & 6.32     & 4  \\
$\rm {MgI_{NLTE}}$ & 7.55     & 3  \\
$\rm {AlI}$    & 6.43         & 2  \\
$\rm {SiI}$    & 7.61         & 12 \\
$\rm {CaI}$    & 6.39         & 16 \\
$\rm {TiI}$    & 4.94         & 33 \\
$\rm {TiII}$   & 4.96         & 12 \\
$\rm {CrI}$    & 5.67         & 32 \\
$\rm {FeI}$    & 7.48         & 145\\
$\rm {FeII}$   & 7.51         & 14 \\
$\rm {NiI}$    & 6.26         & 47 \\
$\rm {BaII}$   & 2.45         & 2  \\
\hline
\end{tabular}
\end{table}
\noindent

With the calibrated linelist, we can obtain refined atmospheric parameters
and abundances for our targets.
Firstly model atmospheres were interpolated from the grid of Kurucz models by
using the values of $\rm {T_{eff}}$, $\rm {log(g)}$, and $\rm {v_{t}}$
determined as explained in the previous section. Then, during the abundance analysis,
$\rm {T_{eff}}$, $\rm {v_{t}}$ and $\rm {log(g)}$ were adjusted in
order to remove trends in Excitation Potential (E.P.) and
equivalent widths vs.\ abundance respectively, and to satisfy
the ionization equilibrium. $\rm {T_{eff}}$ was
optimized by removing any trend in the relation between abundances
obtained from the FeI lines with respect to the E.P.
The optimization of $\rm {log(g)}$ was done in order to satisfy the
ionization equilibrium of species ionized differently: we have used
FeI and FeII lines for this purpose. We changed the value of log(g)
until the following relation was satisfied:

\begin{center}
${\rm log\epsilon(FeII)_{\odot}-log\epsilon(FeI)_{\odot}=log\epsilon(FeII)_{\ast}-log\epsilon(FeI)_{\ast}}$
\end{center}

\noindent
We optimized $\rm {v_{t}}$ by removing any trend in the relation
abundance vs.\ EWs of the spectral lines. This iterative procedure
allowed us to optimize the values of atmospheric parameters on the
basis of spectral data, independent of colors. This is an
important advantage for those clusters, as M4, that are projected in 
regions characterized by a differential reddening.\\
The adopted atmospheric parameters, together with the coordinates, the 
U,B,V,I$_{\rm C}$, and the 2MASS J,H,K magnitudes, are listed in
Tab.~\ref{Tab8}. All the reported magnitudes are not corrected 
for differential reddening.\\
Having $\rm {T_{eff}}$ determinations independent of
colors, we can verify whether the $\rm {T_{eff}}$-scale is affected 
by systematic errors. For this purpose we used the $\rm {T_{eff}}$
and $\rm {[Fe/H]}$ of our stars (see Sec. 6.1) to obtain intrinsic
$\rm {B-V}$ colors
from Alonso's relations. These colors were compared with
WFI photometry corrected for differential reddening.
We obtained a reddening of:

\begin{center}
$\rm E(B-V)=+0.34\pm0.02$
\end{center}

\noindent
This value agrees well with the $\rm {E(B-V)=0.36}$ of Harris (1996)
and with I99 who find $\rm {E(B-V)=0.33\pm0.01}$. Therefore, we
conclude that our $\rm {T_{eff}}$-scale is not affected by strong
systematic errors. \\
Finally we present a $\rm {v_{t}-log(g)}$ relation obtained from our data:

\begin{center}
$\rm {v_t=-(0.254\pm0.016)\cdot log(g) +(1.930\pm0.035)}$
\end{center}

\noindent
It gives  a mean microturbolence velocity 0.15 km/s lower 
than that given by Gratton et al. (1996), but this is not a surprise
because also in other papers (Preston et al. 2006) Gratton et
al. (1996) was found to overestimate the microturbolence, especial in
our $\rm {T_{eff}}$ regime. We underline that our formula is valid 
for objects in the same $\rm {T_{eff}}$-$\rm {log(g)}$ regime as 
out targets, i.e. cold giant stars. 

\subsection{The O synthesis}

Instead of using the EW method, we measured the O content of our stars
by spectral synthesis. This method is necessary because of the blending of the
target O line at 6300 \AA\ with other spectral lines (mainly the Ni transition
at 6300.34 \AA).
For this purpose, we used a linelist from the VALD archive
calibrated on the NOAO Solar spectrum as done before for the other elements,
but in this case we changed the log(gf) parameter of our spectral
lines in order to obtain a good match between our synthetic spectrum
and the observed one.

\begin{figure}
\centering
\includegraphics[bb=18 144 592 475,width=9.6cm]{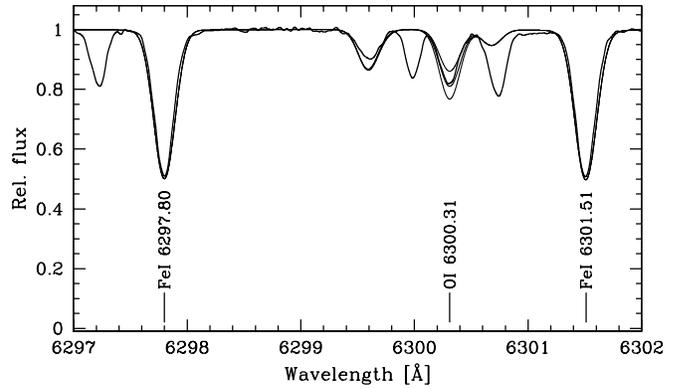}
\caption{Spectrum of the star \#34240, compared with synthetic spectra
  in the region 6297-6302 \AA, which includes the Oxygen line at
  6300.31 \AA. Synthetic spectra were computed for O abundances
  [O/Fe] = +0.30, +0.47, and +0.60 dex. Thick line is the
  observed spectrum, thin lines are the synthetic ones.}
\label{Fig2}
\end{figure}

With the calibrated linelist, it is possible to establish the O
content of our targets. To this aim, we used the standard MOOG running
option $synth$ that computes a set of trial synthetic spectra and
matches these to the observed spectrum. The synthetic spectra were
obtained by using the atmospheric parameters derived in
the previous section. In this way, the Oxygen abundances were deduced
by minimization of the observed-computed spectrum
difference.  An example of a spectral synthesis is plotted in
Fig.~\ref{Fig2} where the spectrum of the observed star \#34240 (thick
line) is compared with synthetic spectra (thin lines) computed for O
abundances $\rm {[O/Fe]}$= +0.30, +0.47, and +0.60 dex. In the plotted
spectral range, the observed spectrum is contaminated by two telluric
features (not present in the models), but none affecting the O line.

The O line is faint, and we could measure the O content only for a
sub-sample of 93 stars. In the remaining 12 spectra, the bad quality of the
Oxygen line prevented us from measuring accurate O abundance.

\begin{figure}
\centering
\includegraphics[bb=18 144 592 518,width=9cm]{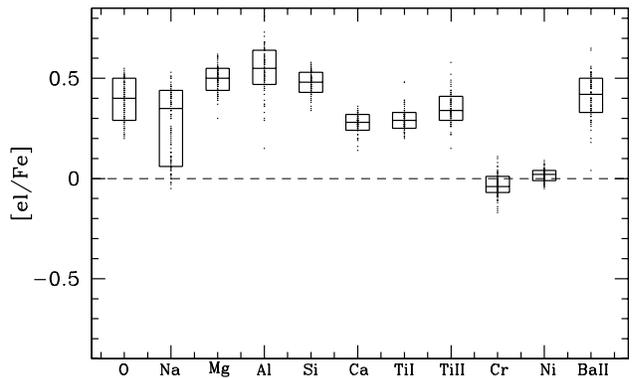}
\caption{Box plot of the M4 giant star element abundances from UVES
         data. For each box the central horizontal line is the median 
         of the data, while the upper and lower lines indicate
         the higher and lower 1$\sigma$ value respectively. 
         The points represent the individual measurements.}
\label{Fig3}
\end{figure}

\section{Results}

The wide spectral range of the UVES data allowed us to derive the
chemical abundances of several elements. The mean abundances
for each element are listed in Tab.~\ref{Tab2} together
with the error of the mean, the rms scatter
and the number of measured stars ($\rm {N_{stars}}$). The rms scatter 
(hereafter $\sigma_{\rm {obs}}$) is assumed to be the $\rm {68.27th}$ percentile
of the distribution of the measures of the single stars and the error 
of the mean is the rms divided by $\sqrt{\rm {N_{stars}-1}}$.
In the last column the abundances derived by I99 were listed as a comparison.

The derived Na and Mg abundances were corrected for the effects of
departures from the LTE assumption (NLTE correction) using the prescriptions by Gratton
et al.\ (1999), as done for the Sun. In the paper all the used Na and Mg abundances are
NLTE corrected, also were not explicitely indicated.
The mean NLTE corrections obtained for [Na/Fe] and [Mg/Fe] are $-0.02$ dex and
$+0.07$ dex, respectively.
For the Fe abundance, we do not distinguish between the results obtained
from I and II ionization stages, because their values are necessarily the same
in the chosen method.

Chemical abundances for the single stars are listed in Tab.~\ref{Tab9}.  
A plot of our measured abundances is shown in Fig.~\ref{Fig3},
where, for each box, the central horizontal line is the median values for the
element, and the upper and lower lines indicate the 1$\sigma$ values higher and lower than the
median values, respectively. The points represent individual measurements.\\

\begin{table} [!htpq]
\caption{The average abundances of M4 stars The results by I99 are
  shown for comparison in Column 5.}
\label{Tab2}
\centering
\begin{tabular}{ l c c c c }
\hline \hline
\multicolumn{4}{c}{This work}                              &  I99    \\
\hline
                        &                 &$\sigma_{\rm {obs}}$ & $\rm {N_{stars}}$ &\\
\hline
${\rm [OI/Fe]}$         & $+0.39\pm0.01$   & 0.09    & 93   & $+0.25\pm0.03$  \\
${\rm [NaI/Fe]_{NLTE}}$ & $+0.27\pm0.02$   & 0.17    & 105  & $+0.22\pm0.05$  \\
${\rm [MgI/Fe]_{NLTE}}$ & $+0.50\pm0.01$   & 0.06    & 105  & $+0.44\pm0.02$  \\
${\rm [AlI/Fe]}$        & $+0.54\pm0.01$   & 0.11    & 87   & $+0.64\pm0.03$   \\
${\rm [SiI/Fe]}$        & $+0.48\pm0.01$   & 0.05    & 105  & $+0.55\pm0.02$   \\
${\rm [CaI/Fe]}$        & $+0.28\pm0.01$   & 0.04    & 105  & $+0.26\pm0.02$   \\
${\rm [TiI/Fe]}$        & $+0.29\pm0.01$   & 0.05    & 105  & $+0.30\pm0.01$   \\
${\rm [TiII/Fe]}$       & $+0.35\pm0.01$   & 0.06    & 105  & $+0.30\pm0.01$  \\
${\rm [CrI/Fe]}$        & $-0.04\pm0.01$   & 0.05    & 105  &     --           \\
${\rm [FeI/H]}$         & $-1.07\pm0.01$   & 0.05    & 105  & $-1.18\pm0.00$   \\
${\rm [NiI/Fe]}$        & $+0.02\pm0.01$   & 0.03    & 105  & $+0.05\pm0.01$   \\
${\rm [BaII/Fe]}$       & $+0.41\pm0.01$   & 0.09    & 103  & $+0.60\pm0.02$   \\
\hline
\end{tabular}
\end{table}

\subsection{Internal errors associated to the chemical abundances}

\noindent The measured abundances of of every element vary from
star to star as a consequence of both measurement errors and
intrinsic star to star abundance variations.
In this section, our final goal is to search for evidence of intrinsic
abundance dispersion in each element by comparing the observed
dispersion $\sigma_{\rm {obs}}$ and that produced by internal errors
($\sigma_{\rm {tot}}$). Clearly, this requires an accurate analysis of
all the internal sources of measurement errors.
We remark here that we are interested in star-to-star intrinsic
abundance variation, i.e. we want to measure the internal intrinsic
abundance spread of our sample of stars. For this reason, we
are not interested in external sources of error which are systematic
and do not affect relative abundances.\\
It must be noted that two sources of errors mainly contribute
to $\sigma_{\rm {tot}}$. They are:
\begin{itemize}
\item the errors $\sigma_{\rm {EW}}$ due to the uncertainties in the
  EWs measures;
\item the uncertainty $\sigma_{\rm {atm}}$ introduced by errors in the 
  atmospheric parameters adopted to compute the chemical abundances.
\end{itemize}

\noindent In order to derive an estimate of $\sigma_{EW}$, we consider that
EWs of spectral lines in stars with the same atmospherical
parameters and abundances are expected to be equal; any difference in the
observed EWs can be attributed to measurements errors. Hence, in order
to estimate $\sigma_{\rm {EW}}$, we applied the following procedure:
\begin{itemize}
\item{Select two stars (\#33683 and \#33946) from our sample
  characterized by exactly the same $\rm {T_{eff}}$, and $\rm {log(g)}$, $\rm {v_{t}}$,
  [Fe/H] within a range of 0.05;}
\item {Compare the EWs of the iron spectral lines in these two stars
  and calculate the standard deviation of the distribution of the
  differences between the EWs divided by $\sqrt{2}$. We found a value
  of 3.6 m\AA \ that 
  we assume to represent our estimate for the errors in the EWs.
  Iron lines were selected because Fe abundance has always been found to be
  the same for stars in GCs, with the exception of $\omega$~Cen. This
  means that any differences in the EW for the same couple of lines is
  due only to measurement errors;}  
\item {Calculate the corresponding error in abundance measurements
  ($\sigma_{\rm {EW}}$) by using a star (\#21728) at intermediate
  temperature, assumed to be representative of the entire sample.
  To this aim, we selected two spectral lines for each element
  in order to cover the whole E.P. range,
  changed the EWs by 3.6 m\AA \ and calculated the corresponding mean
  chemical abundance variation. This number divided by $\sqrt {\rm
    {N_{lines}-1}}$ is our $\sigma_{\rm {EW}}$. Results are listed in
  Tab.~\ref{Tab3}}. 
\end{itemize}

\noindent A  more elaborate analysis is required to determine
$\sigma_{\rm {atm}}$.
We followed two different approaches.

In order to better understand the method, it is important
to summarize the role of the atmospheric parameters in the determination
of chemical abundances.
As fully described in sections 4.1, the best estimate of chemical
abundances, $\rm {T_{eff}}$, $\rm {v_{t}}$ and $\rm {log(g)}$ obtained
from the spectrum of a single star are those that satisfy at the same
time three conditions:
\begin{itemize}
\item {removing any trend from the straight line that best fits the
  abundances vs.\ E.P.;}
\item {removing any trend from the straight line that best fits the
  abundances as a function of equivalent widths;}
\item {satisfy the ionization equilibrium through the condition $\rm
  {[FeI/H]=[FeII/H]}$.} 
\end{itemize}
Therefore, any error in the slope of the best fitting lines 
and in FeI/FeII determinations produces an error on the measured abundance.

In order to derive the error in temperature we applied the
following procedure.
First we calculated, for each star, the errors associated with the slopes of
the best least squares fit in the relations between 
abundance vs.\ E.P.
The average of the errors corresponds to the typical error on the slope.
Then, we selected three stars representative of the entire sample
(\#29545, \#21728, and \#34006) with high, intermediate, and low
$\rm {T_{eff}}$, respectively.
For each of them, we fixed the other parameters and varied the temperature
until the slope of the line that best fits the relation between
abundances and E.P. became equal to the respective
mean error. This difference in temperature represents an estimate of
the error in temperature itself. The value we found is $\rm {\Delta
  T_{\rm eff} = 40}$ K. 

\begin{figure*}[!htpb]
\centering
\includegraphics[bb=170 385 450 590,width=6.5cm]{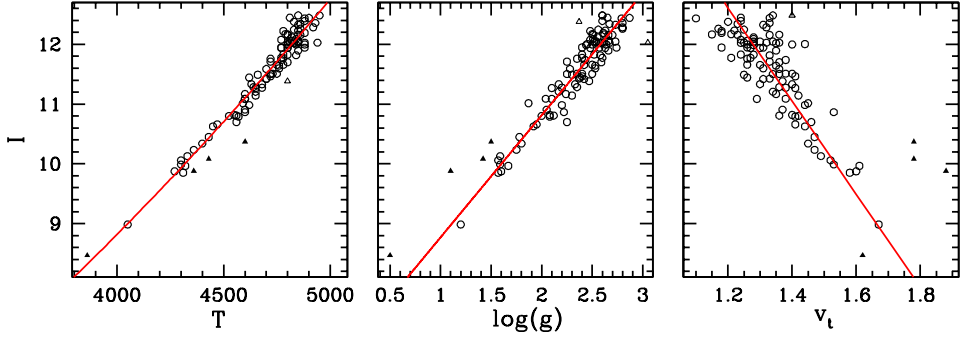}
\caption{$\rm {I}$ magnitude (corrected for differential reddening) as
  a function of the atmospheric parameters for our target stars. For
  $\rm {T_{eff}}$ the data were fitted with a parabolic curve, for
  log(g) and $\rm {v_{t}}$ with a straight line.} 
\label{Fig4}
\end{figure*}

The same procedure was applied for $\rm {v_{t}}$, but using
the relation between abundance and EWs. We obtained a
mean error of $\Delta\rm {v_{t}}=0.06$ km/s.

Since $\rm {log(g)}$ has been obtained by imposing the condition $\rm
{[FeI/H]=[FeII/H]}$, and the measures of $\rm {[FeI/H]}$ and $\rm
{[FeII/H]}$ have averaged uncertainties of 
$\overline {\sigma_{\rm {star}}[\rm {FeII/H}]}$ and $\overline {\sigma_{\rm {star}}[\rm
{FeI/H}]}$ (where $\sigma_{\rm {star}}[\rm {Fe/H}]$ is the dispersion of
the iron abundances derived by the various spectral lines in
each spectrum and given by MOOG, divided by $\sqrt {\rm {N_{lines}-1}}$), in order to
associate an error to the measures of gravity we have varied the
gravity of the three representative stars such that the relation:
\begin{center}
$\rm {[FeI/H] - \overline{\sigma_{star}[FeI/H]} = [FeII/H] + \overline{\sigma_{star}[FeII/H]}}$
\end{center}
was satisfied. The obtained mean error is $\Delta [\rm {log(g)}]$=0.12.

Once the internal errors associated to the atmospheric parameters were
calculated, we re-derived the abundances of the three reference stars by
using the following combination of atmospheric parameters:
\begin{itemize}
\item ($\rm {T_{eff}} \pm \Delta (\rm {T_{eff}})$, $\rm {log(g)}$,  $\rm {v_{t}}$)
\item ($\rm {T_{eff}} $, $\rm {log(g)} \pm \Delta (\rm {log(g)})$,  $\rm {v_{t}}$)
\item ($\rm {T_{eff}} $, $\rm {log(g)}$,  $\rm {v_{t}} \pm \Delta (\rm {v_{t}})$)
\end{itemize}
where $\rm {T_{eff}}$, $\rm {log(g)}$,  $\rm {v_{t}}$ are the measures
determined in Section 4.2.

The resulting errors in the chemical abundances due to uncertainties in
each atmospheric parameter are listed in Tab.~\ref{Tab3} (columns 2, 3
and 4). The values of $\sigma_{\rm {atm}}$ (given by the squared sum of
the uncertainties introduced by each single parameter) listed in
column 5 are our final estimates of the error introduced by the
uncertainties of all the atmospheric parameters on the chemical
abundance measurements. 

We also used a second approach to estimate $\sigma_{\rm {atm}}$ as
confirmation of the first method. In order
to verify our derived uncertainties related to the temperature,
gravity and microturbolence, we considered that RGB stars with the
same $\rm {I}$ magnitude must have the same atmospheric parameters.
Hence, we started by plotting the magnitude $\rm {I}$ as a function of
$\rm {T_{eff}}$, $\rm {log(g)}$ and $\rm {v_{t}}$ as shown in
Fig.~\ref{Fig4}. We see in this figure that the typical internal error in our
$\rm {I}$ photometry ($\sim0.01$ mag) translates into an error in temperature
of $\sim3$ K, in $\rm {log(g)}$ of $\sim0.005$, and in $\rm {v_{t}}$ of $\sim0.002$ km/s, each of which is
absolutely negligible. The data were fitted by a parabolic curve in the case of $\rm {T_{eff}}$,
and by straight lines in the cases of $\rm {log(g)}$ and $\rm
{v_{t}}$. We determined the differences between the $\rm {T_{eff}}$,
log(g) and $\rm {v_{t}}$ of each star and the corresponding value on the fitting
curve. Assuming the 68.27th percentile of the absolute values of these
differences as an estimate of the dispersion of the points
around the fitting curve, all the stars with a distance from the curve
larger than $3\sigma$ were rejected (the open triangles in
Fig.~\ref{Fig4}). The four probable AGB stars (filled triangles) that
are evident on the $\rm {B-(B-I)}$ CMD of Fig.~\ref{Fig1} were also rejected.
With the remaining stars, the curve was refitted and the differences
between the atmospheric parameters of the stars and the fit were
redetermined. The 68.27th percentile of their absolute
values are our estimate of the uncertainties on the atmospheric
parameters.

By using this second method we obtained the uncertainties $\Delta {\rm
  {T_{eff}}}=$37 K, $\Delta \rm {log(g)}=$0.12 and $\Delta \rm {v_{t}}=$0.06
km/s, consistent with the ones determined with the first method.

\begin{table*}[!htpq]
\caption{Sensitivity of derived UVES abundances to the atmospheric parameters
  and EWs. We reported the total error due to the atmospheric parameters
  ($\sigma_{\rm atm}$), due to the error in EW measurement
  ($\sigma_{\rm EW}$), the squared sum of the two ($\sigma_{\rm tot}$), and the
  observed dispersion ($\sigma_{\rm obs}$) for each element. The FeII
  observed dispersion is not reported because it is necessarily the
  same as FeI due to the method used in this paper to obtain
  atmospheric parameters.} 
\centering
\label{Tab3}
\begin{tabular}{ l r r r c c c c c}
\hline\hline
\ & $\rm {\Delta T_{eff}} (K)$ & $\Delta$log(g) & $\Delta \rm {v_{t}}$ (km/s) & \
$\sigma_{\rm {atm}}$ & $\sigma_{\rm {EW}}$ & $\sigma_{\rm {tot}}$ & $\sigma_{\rm {obs}}$\\
\hline
                      &$+40$        &$+0.12$ &$+0.06$&      &       &      &   \\ \hline
 ${\rm [OI/Fe]}$      &--           &--      &--     & --   & --    & 0.04 &0.09   \\
 ${\rm [NaI/Fe]}$     &$+0.00$      &$-0.01$ &$+0.02$& 0.02 & 0.04  & 0.04 &0.17   \\
 ${\rm [MgI/Fe]}$     &$-0.01$      &$-0.01$ &$+0.01$& 0.02 & 0.06  & 0.06 &0.06   \\
 ${\rm [AlI/Fe]}$     &$-0.01$      &$-0.02$ &$+0.02$& 0.03 & 0.06  & 0.07 &0.11   \\
 ${\rm [SiI/Fe]}$     &$-0.04$      &$+0.03$ &$+0.02$& 0.05 & 0.03  & 0.06 &0.05   \\
 ${\rm [CaI/Fe]}$     &$+0.01$      &$-0.02$ &$+0.00$& 0.02 & 0.02  & 0.03 &0.04   \\
 ${\rm [TiI/Fe]}$     &$+0.04$      &$-0.02$ &$+0.00$& 0.04 & 0.02  & 0.04 &0.05   \\
 ${\rm [TiII/Fe]}$    &$+0.03$      &$-0.01$ &$-0.01$& 0.03 & 0.05  & 0.06 &0.06   \\
 ${\rm [CrI/Fe]}$     &$+0.02$      &$-0.01$ &$+0.00$& 0.02 & 0.06  & 0.06 &0.05   \\
 ${\rm [FeI/H]}$      &$+0.04$      &$+0.00$ &$-0.03$& 0.05 & 0.01  & 0.05 &0.05   \\
 ${\rm [FeII/H]}$     &$-0.03$      &$+0.06$ &$-0.01$& 0.07 & 0.04  & 0.08 &--    \\
 ${\rm [NiI/Fe]}$     &$-0.01$      &$+0.02$ &$+0.01$& 0.02 & 0.02  & 0.03 &0.03   \\
 ${\rm [BaII/Fe]}$    &$+0.05$      &$-0.04$ &$-0.03$& 0.07 & 0.05  & 0.09 &0.09   \\
\hline
\end{tabular}
\end{table*}

Our best estimate of the total error associated to the abundance
measures is calculated as
\begin{center}
$\rm {\sigma_{tot}=\sqrt{\sigma_{EW}^{2}+\sigma_{atm}^{2}}}$
\end{center}
listed in the column 7 of Tab.~\ref{Tab3}.
For O $\rm {\sigma_{tot}}$ was obtained in a different way since its measure
was not based on the EW method (see Section 6.4). In all the plots for the
error bars associated with the measure of abundance we adopted $\sigma_{\rm tot}$.  

In addition, systematic errors on abundances can be introduced by
systematic errors in the scales of
$\rm {T_{eff}}$, log(g), $\rm {v_{t}}$, and by deviations of the real
stellar atmosphere from the model (i.e. deviations from LTE).
A study of the systematic effects goes beyond the purposes of
this study, where we are more interested in internal star to star variation in
chemical abundances.

Comparing $\sigma_{\rm {tot}}$ with the observed
dispersion $\sigma_{\rm {obs}}$ (column 8) at least for the abundance
ratios ${\rm [Na/Fe]}$, ${\rm [O/Fe]}$ and $\rm {[Al/Fe]}$, we found
$\sigma_{\rm {obs}}$ significantly larger than $\sigma_{\rm {tot}}$,
suggesting the presence of a real spread in the content of these
elements within the cluster. 

\section{The chemical composition of M4}
In the following sections, we will discuss in detail the measured
chemical abundances. In addition, we will use the abundances of C and
N from the literature to extend our analysis to those elements.
The errors we give here are only internal. External ones can be estimated
by comparison with other works (see Tab.~\ref{Tab2}). 

\subsection{Iron and iron-peak elements}

The weighted mean of the [Fe/H] found in the 105 cluster members is

\begin{center}
${\rm [Fe/H]=-1.07\pm0.01\ dex}$
\end{center}

\noindent
The other Fe-peak elements (Ni and Cr) show roughly the Solar-scaled
abundance.

The agreement between $\sigma_{\rm obs}$ and $\sigma_{\rm {tot}}$ (see
Tab.~\ref{Tab3}), allows us to conclude that there 
is no evidence for an internal dispersion in the iron-peak 
element content, suggesting that this cluster is homogeneous in metallicity
down to the $\sigma_{\rm {obs}}$ level.

\subsection{$\alpha$ elements}

\noindent The chemical abundances for the $\alpha$ elements
O, Mg, Si, Ca, and Ti are listed in Tab.~\ref{Tab2}. All
the $\alpha$ elements are overabundant. Ca and Ti are
enhanced by $\sim$0.3 dex, while Si and Mg by $\sim$0.5 dex.
The results for $\rm {[O/Fe]}$ will be discussed in detail in Section 6.4.

Previous investigators (GQO86, BW92 and I99) have also found significant
overabundances for both Si and Mg. In particular, for $\rm {[Si/Fe]}$ they found
values higher than 0.5 dex, slightly higher than our results, while our
$\rm {[Mg/Fe]} = 0.50$ lies in the middle of
the literature results that range between 0.37 (BW92) and 0.68 dex (GQO86).
As noted by I99, the abundances of these two elements in M4 are higher if
compared to the ones found in other globular clusters showing a similar
metallicity, i. e. M5 (Ivans et al.\ 2000). I99 suggest that the higher
abundances in M4 should be primordial, e.g. due to the chemical composition
of the primordial site from which the cluster formed.

For $\rm {[Ca/Fe]}$ we obtained a scatter of 0.04 dex,
differing from I99, who found $\sigma$=0.11 dex (see discussion in
Section $6.6$).
TiI and TiII show the same abundance within $3\sigma$ (see Tab.~\ref{Tab2}),
with a  mean of the two different ionized species of
$\rm {[Ti/Fe]}=+0.32\pm0.04$.

The average of all the $\alpha$ elements, gives an $\alpha$
enhancement for M4 of:

\begin{center}
${\rm [\alpha/Fe]=+0.39\pm0.05 \ dex}$
\end{center}

\noindent
In this case the agreement between $\sigma_{\rm obs}$
and $\sigma_{\rm {tot}}$ again allows us to conclude that, down to the
$\sigma_{\rm obs}$ level, there 
is no evidence of internal
dispersion for these elements, with the exception of O, which we will discuss in Section 6.4.

\subsection{Barium}

As in I99 and BW92, we have found a strong overabundance of Ba. We
have $\rm {[Ba/Fe]}=0.41\pm 0.01$ which is about 0.2 dex smaller than
in I99. In any case, our results confirm that Ba is significantly
overabundant in M4, at variance with what found by GQO86, possibly
solving this important issue.  As deeply discussed by I99, this high
Barium content cannot be accounted for mixing processes, but must be a
signature of the presence of s-process elements in the primordial
material from which M4 stars formed. The excess of s-process elements
provides some support to the idea 
that the formation of the stars we now observe in
M4 happened after intermediate-mass AGB stars have polluted the
environment, or that it lasted for long enough to allow intermediate-mass AGB
stars to strongly pollute the lower mass forming stars. We
do not find any significant dispersion in Ba content for the bulk of
our target stars, though there are a few outliers worth further discussion (see below). 

\subsection{Na-O anticorrelation}

Variations in light element abundances is common among GCs,
and is also present in M4.
In particular, Sodium and Oxygen have
very large dispersions: we obtained $\sigma_{\rm {\rm [Na/Fe]}}$=0.17
and $\sigma_{\rm [O/Fe]}$=0.09.
A large star to star scatter in Sodium and Oxygen abundance in M4 was found
also in other previous studies (I99, D92).

\begin{figure}[!hpb]
\centering
\includegraphics[width=9cm]{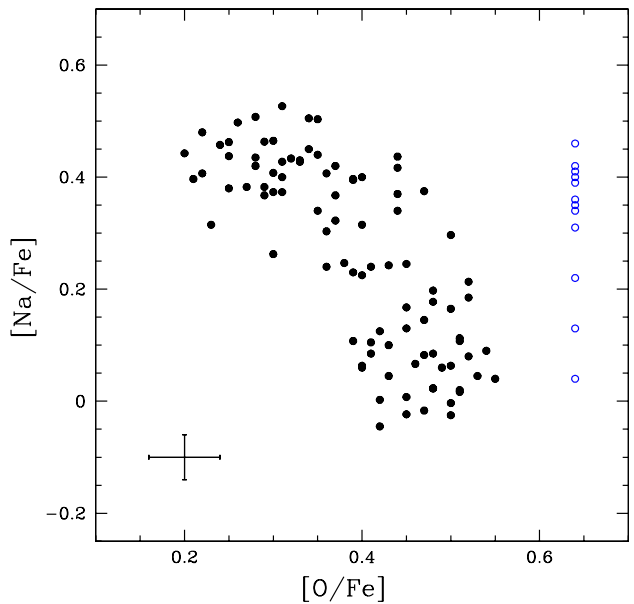}
\caption{${\rm [Na/Fe]}$ vs.\ $\rm {[O/Fe]}$ abundance ratios. The error
  bars represent the typical errors $\sigma_{\rm tot}$ from Tab.~\ref{Tab3}.}
\label{Fig5}
\end{figure}

\begin{figure}[!hpb]
\centering
\includegraphics[width=8.5cm]{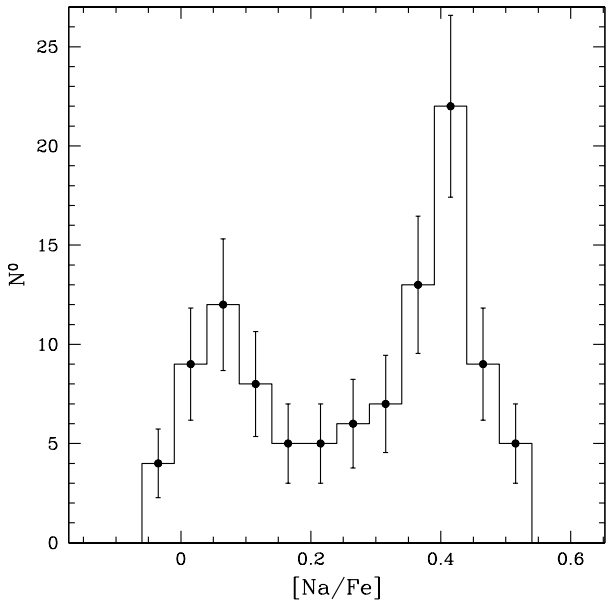}
\caption{Histogram of the distribution of the ${\rm [Na/Fe]}$
  abundances. The error bars represent the Poisson errors.}
\label{Fig6}
\end{figure}

Sodium and Oxygen show the typical anticorrelation found in many other
GCs (Gratton et al.\ 2004), as shown in Fig.~\ref{Fig5}, where the ${\rm
[Na/Fe]}$ values are plotted as a function of ${\rm [O/Fe]}$.  The
open blue circles represent the 12 stars for which it was not possible to obtain
a good estimate of the Oxygen abundance.

Since the Oxygen abundance was derived from just one spectral line, we adopted
the following the following procedure to derive a raw estimate of the
typical error $\sigma_{\rm {tot}}$ associated with each measure: we selected the group of
stars with ${\rm [Na/Fe]}$$<$0.20, assuming them to be homogeneous in
their O content. Under this hypothesis, the dispersion in O can be
assumed to be due to the random error associated with the measured O
abundance. We obtained $\sigma_{\rm [O/Fe]}=0.04$ dex, which
corresponds to the error bar size in Fig.~\ref{Fig5} and to the
$\sigma_{\rm tot}$ in Tab.~\ref{Tab3}. 

The only previous study showing the Na-O anticorrelation in M4 was that by
I99. They analysed ${\rm [Na/Fe]}$ and ${\rm [O/Fe]}$ abundances
for 24 giant stars from high resolution spectra taken at the Lick and
McDonald Observatories. With respect to their work, our dispersion for
[O/Fe] is quite a bit smaller (0.09 dex against their value of 0.14 dex),
and our average value is larger (see Tab.~\ref{Tab2}).
We have seven stars in common with their high resolution sample, and another
seven with their medium resolution sample for which they derived only
the Oxygen abundance. The
comparison between their results and the present ones are listed in
Tab.~\ref{Tab4}, where the last two columns give
the differences between I99 and our results. We can note that our
abundances are systematically higher than those of I99. The effect is
higher for the ${\rm [O/Fe]}$, but is present for Sodium too, and could
reflect the difference of $\sim$ 0.1 dex we found for the iron with
respect to I99.
This systematic difference does not affect the
shape of the Na-O anticorrelation, but it does emphasize the
overabundance of O in M4 stars with respect to other clusters with
similar metallicity, as already noticed by I99: M4 stars formed in an
environment particularly rich in O and possibly in Na.

The most interesting result of our investigation is that,
thanks to the large number of stars in our sample, we can show that
the ${\rm [Na/Fe]}$ distribution is bimodal (Fig.~\ref{Fig6}).
Setting an arbitrary separation between the two
peaks in Fig.~\ref{Fig6} at $\sim$0.2 dex, we obtain, for the two
groups of stars with higher and lower Na content, a mean ${\rm
[Na/Fe]}=0.38\pm0.01$ and ${\rm [Na/Fe]}=0.07\pm0.01$,
respectively.

\begin{figure}[!hpb]
\centering
\includegraphics[width=8.7cm]{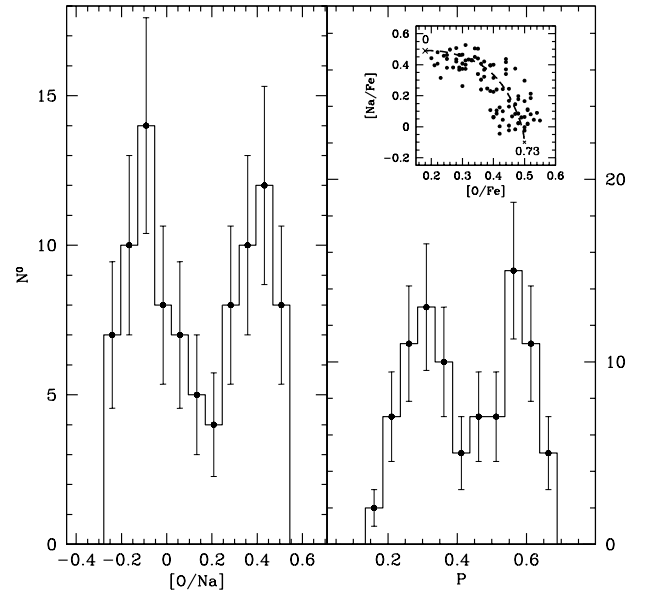}
\caption{Left: Distribution of stars along the Na-O anticorrelation
         represented by the $\rm {[O/Na]}$ ratios.
         Right: Distribution of the projected position P of stars
         on the parametric curve plotted in the inset pannel. The
         coordinates of the edges of the curve are indicated.}
\label{Fig7}
\end{figure}

\begin{table*}[!htpq]
\caption{Comparison between our ${\rm [Na/Fe]}$ and $\rm {[O/Fe]}$ and
  those of I99 for the stars in common. The Oxygen values were
  compared both with 
  the high resolution sample of I99 ($\rm {[O/Fe]_{I99}^{(a)}}$) and
  with the medium resolution sample ($\rm {[O/Fe]_{I99}^{(b)}}$). The
  last three columns give the differences $\rm {[el/Fe]_{this~work}-[el/Fe]_{I99}}$.}
\centering
\label{Tab4}
\begin{tabular}{ c c c c c c c c c }
\hline
\hline
ID   &${\rm [OI/Fe]_{this~work}}$&$\rm {[OI/Fe]_{I99}^{(a)}}$&$\rm {[OI/Fe]_{\rm I99}^{(b)}}$&$\rm {[NaI/Fe]_{this~work}}$&$\rm {[NaI/Fe]_{I99}^{(a)}}$&$\Delta({\rm O^{(a)}})$&$\Delta({\rm O^{(b)}})$&$\Delta({\rm Na})$\\
\hline
L1411  &$+0.28$  &$+0.20$  &$+0.07$&$+0.51$ &$+0.43$&$+0.08$  &$+0.21$   &$+0.08$  \\
L1501  &$+0.34$  &$+0.10$  &$ --  $&$+0.51$ &$+0.42$&$+0.24$  &$--$      &$+0.09$  \\
L1514  &$+0.50$  &$+0.41$  &$+0.41$&$+0.17$ &$+0.01$&$+0.09$  &$+0.09$   &$+0.16$  \\
L2519  &$+0.48$  &$+0.37$  &$+0.37$&$+0.02$ &$-0.19$&$+0.11$  &$+0.11$   &$+0.21$  \\
L2617  &$+0.25$  &$+0.01$  &$+0.07$&$+0.44$ &$+0.50$&$+0.24$  &$+0.18$   &$-0.06$  \\
L3612  &$+0.25$  &$+0.10$  &$+0.12$&$+0.46$ &$+0.47$&$+0.15$  &$+0.13$   &$-0.01$  \\
L3624  &$+0.48$  &$+0.29$  &$+0.27$&$+0.18$ &$+0.10$&$+0.19$  &$+0.21$   &$+0.08$  \\
L1403  &$+0.31$  &$--$     &$+0.22$&$+0.53$ &$--$   &$--$     &$+0.09$   &$--$ \\
L2410  &$+0.39$  &$--$     &$+0.32$&$+0.11$ &$--$   &$--$     &$+0.07$   &$--$ \\
L2608  &$+0.43$  &$--$     &$+0.37$&$+0.24$ &$--$   &$--$     &$+0.06$   &$--$ \\
L1617  &$+0.24$  &$--$     &$+0.17$&$+0.46$ &$--$   &$--$     &$+0.07$   &$--$ \\
L4415  &$+0.44$  &$--$     &$+0.37$&$+0.37$ &$--$   &$--$     &$+0.07$   &$--$ \\
L4421  &$+0.28$  &$--$     &$+0.07$&$+0.44$ &$--$   &$--$     &$+0.21$   &$--$ \\
L4413  &$+0.22$  &$--$     &$+0.27$&$+0.48$ &$--$   &$--$     &$-0.05$   &$--$ \\
\hline
mean   &$+0.35$  &$+0.21$  &$+0.24$&$+0.35$ &$+0.25$&  $+0.16$&  $+0.11$&$+0.08$\\
%$\pm$  &$~~0.03$ &$~~0.06$ &$~~0.04$&$~~0.05$&$~~0.11$&$~~0.03$&$~~0.02$&$~~0.04$\\
\hline
\end{tabular}
\end{table*}

In correspondence with the two peaks in the ${\rm [Na/Fe]}$
distribution, there are also two $\rm {[O/Fe]}$ groups
(Fig.~\ref{Fig5}): the first group is centered at $\rm {[O/Fe]\sim0.30}$,
while the second is at $\rm {[O/Fe]\sim0.47}$, with a few stars
having intermediate Oxygen abundance.\\

We have followed two methods to trace the distribution of the stars on the Na-O
plane.
The first method uses the ratio $\rm {[O/Na]}$ between the O and the Na
abundances (Carretta et al.\ 2006) which appears to be the
best indicator to trace the star distribution along the Na-O
anticorrelation, because this ratio continues to vary even at the
extreme values along the distribution. The distribution of
the [O/Na] is plotted in Fig.~\ref{Fig7} (left panel).
We can identify two bulks of RGB stars in this plot: one
at $\rm {[O/Na]\sim0.40}$ and the second one at $\rm {[O/Na]\sim-0.10}$.
There might be another group of stars between
[O/Na]$\sim$0 and [O/Na]$\sim$0.25, but we cannot provide conclusive
evidence.

Note that the two groups of stars in the Na-O anticorrelation have
their corresponding peaks at $\rm{[O/Fe]\sim0.47}$ and $\rm {[O/Fe]\sim0.30}$,
so both groups are O-enhanced.\\
There are no differences in $\rm {[Fe/H]}$ content between these
two groups of stars: both of them have $\rm {[Fe/H]=-1.07}$ dex.

We have also analysed the distribution of stars along the Na-O
anticorrelation with an alternative procedure whose steps are
briefly described below:

\begin{itemize}

\item{A parametric curve following the observed Na-O distribution as shown
  in the inset of the right panel of Fig.~\ref{Fig7} has been traced;}

\item{Each observed point in the Na-O distribution has been projected
  on the parametric curve;}

\item For each projected point, the distance (P) from the origin of
the parametric curve (indicated with 0 in the inset of the
right panel of Fig.~\ref{Fig7}) has been calculated;

\item{A histogram of the P distance distribution has been constructed.}

\end{itemize}
The histogram is shown in the right panel of Fig.~\ref{Fig7}.
In this case also, two peaks are evident: the
first one at $\rm {P\sim0.28}$, and the second one at $\rm {P\sim0.58}$.

In conclusion, we obtained the Na-O anticorrelation for a large sample
(93 RGB stars) in the globular cluster M4.
The distribution of the objects on the Na-O anticorrelation
is clearly bimodal.

\subsection{Mg-Al anticorrelation}

\begin{figure}[!hpb]
\centering
\includegraphics[width=9cm]{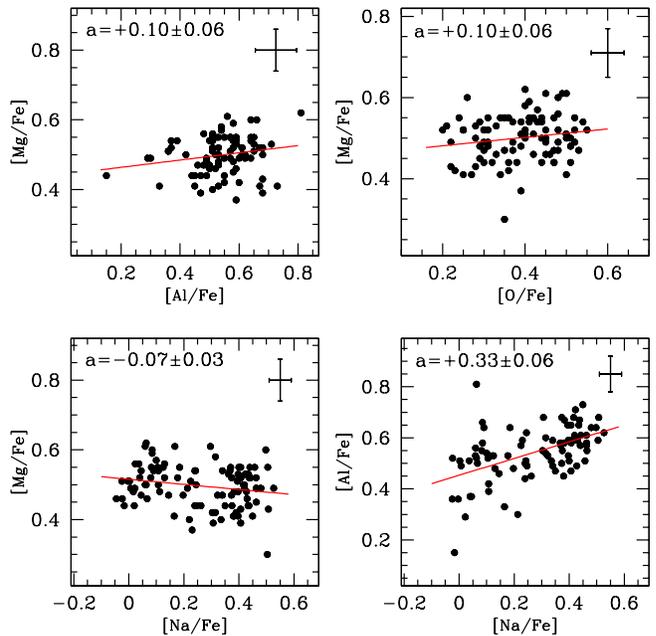}
\caption{Bottom panels: $\rm {[Al/Fe]}$ and $\rm {[Mg/Fe]}$ abundance ratios
         are plotted as a function of ${\rm [Na/Fe]}$. Upper panels:
         $\rm {[Mg/Fe]}$ ratio as a function of $\rm {[O/Fe]}$ and
         $\rm {[Al/Fe]}$. In each panel the red line is the  best
         least squares fit and $\rm {a}$ is the slope of this
         line. The error bars represent the typical errors
         $\sigma_{\rm tot}$ from Tab.~\ref{Tab3}.} 
\label{Fig8}
\end{figure}

Aluminium abundances have been determined from the EWs of the AlI lines at
6696 \AA \ and 6698 \AA, and Mg abundances from the EWs of the MgI
doublet at 6319 \AA, 6318 \AA , and of the line at
5711\AA. Figure~\ref{Fig8} shows the [Mg/Fe] as a function of
[Al/Fe] and [O/Fe] (upper panels) and [Mg/Fe] and [Al/Fe] as a
function of [Na/Fe] (lower panels).

In each panel the red line represents the best least squares fit to the data.
The value $\rm {a}$ is the slope of the best fit straight line $\rm {y=ax+b}$.
There is no clear Mg-Al anticorrelation (upper left panel of
Fig.~\ref{Fig8}), although Al is more spread out than Mg ($\rm
{\sigma_{[Al/Fe]}=0.11}$ vs.\ $\rm {\sigma_{[Mg/Fe]}=0.06}$). Both Al and
Mg are overabundant, with average value of $\rm {[Al/Fe]=0.54\pm0.01}$
and $\rm {[Mg/Fe]=0.50\pm0.01}$ (internal errors).

The right-bottom panel of Fig.~\ref{Fig8} shows that there is a
correlation between $[\rm {Al/Fe]}$ and $\rm {[Na/Fe]}$ (higher Al
abundances for higher Na content), though it is less pronounced than
the Na-O anticorrelation. Aluminium spans over a range of $\sim0.4$ dex,
while Sodium covers a range of $\sim0.6$ dex. The spread in Al content of M4
is smaller than in other clusters of similar metallicity, like M5
(Ivans et al.\ 1999) and M13 (Johnson et al.\ 2005).

\begin{figure}[!hpbt]
\centering
\includegraphics*[width=8.cm]{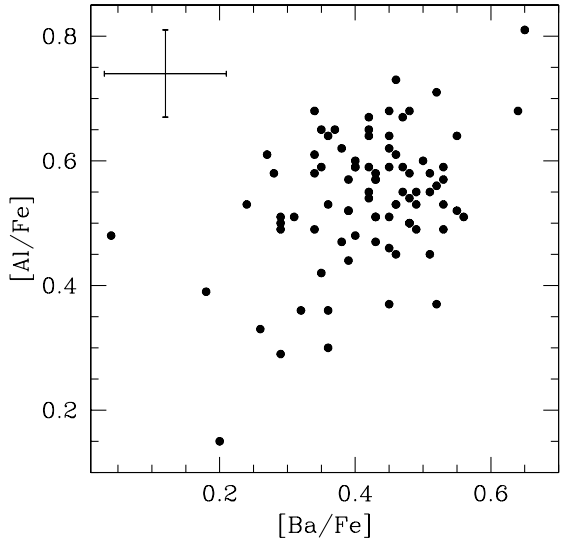}
\caption{$\rm {[Al/Fe]}$ vs.\ $\rm {[Ba/Fe]}$ abundance ratios. The
  error bars represent the typical errors $\sigma_{\rm tot}$ from
  Tab.~\ref{Tab3}.}   
\label{Fig9}
\end{figure}

\noindent Figure~\ref{Fig9} shows that there might also be a
correlation between the ${\rm [Al/Fe]}$ and ${\rm [Ba/Fe]}$.

Mg does not clearly correlate neither with Na (left-bottom panel of
Fig.~\ref{Fig8}) nor O (upper-right panel of Fig.~\ref{Fig8}).

The absence of a Mg-Al anticorrelation is somehow surprising, in view of the
presence of a strong double peaked
Na-O anticorrelation, and of a correlation between Al and
Na and possibly Al and Ba. All these correlations seems to indicate the
presence of material that has gone through the s-process phenomenon.
Assuming that the Na enhancement comes from the proton capture
process at the expense of Ne, we can also expect that a similar process is
the basis of the Al enhancement at the expense of Mg. Still, we do
not find any significant dispersion in the Mg content. This problem has
already  been noted by I99, who suggested that the required Mg destruction
is too low to be measured (see their Section 4.2.2 for further details).

Here we only add that, despite our large sample,
neither in the Al nor in the Mg (and Ba) distribution
can we see any evidence of the two peaks so clearly visible as in the
Na and O distribution.
But, in spite of the lack of a clear Mg-Al anticorreletion, we found that Na
poor stars have on average higher Mg and lower Al. The difference
between the median Mg and Al abundances for the Na-poor and
Na-rich samples are: $\rm {[Mg/Fe]_{Na-rich}-[Mg/Fe]_{Na-poor}=-0.04\pm0.01}$ and
$\rm {[Al/Fe]_{Na-rich}-[Al/Fe]_{Na-poor}=+0.08\pm0.02}$
This should be consistent with the scenario proposed by I99 (see their
Sec. 4.2.2 for more details) who predicted that a drop of only 0.05
dex in Mg is needed to account for the observed increase in the abundance of Al.
(Obviously this requires the further hypothesis that all the
Na enhanced  stars are also Al enhanced.)

\subsection{CN bimodality}

In this section we discuss the correlation of the
$\lambda$3883 CN absorption band strength with Na, O, and Al abundances.
Previous studies revealed a trend of Na and Al with the CN band
strength. By studying 4 RGB stars from high
resolution spectra,
D92 found higher Sodium abundances in the two CN-strong stars with
respect to the two CN-weak ones (Suntzeff \& Smith 1991).
I99 found that the Oxygen abundance is anticorrelated
with Nitrogen, whereas both Na and Al are more abundant in CN-strong
giants than in CN-weak ones.

In our study, two peaks in the Na distribution are evident
(see Fig.~\ref{Fig6}), while there is no equally strong
pattern in the Al distribution (right bottom panel of
Fig.~\ref{Fig8}).
The CN band strengths of red giants in M4 have been measured by
N81, Suntzeff \& Smith (1991), and I99. 
Smith \& Briley 2005 (SB05) have homogenized all the available data for
CN band strengths in terms of the S(3839) index, e. g. the
ratio of the flux intensities of the Cyanogen band near 3839 \AA \ and the
nearby continuum.
In the following, we will use the S(3839) index calculated by SB05.

In Tab.~\ref{Tab5} the CN-index S(3839), the $\rm {[C/Fe]}$ and the
$\rm {[N/Fe]}$ values (taken from SB05 and I99, respectively), and
the identification as CN-strong (S), -weak (W), and -intermediate (I),
are listed for our target stars with available CN data (34 stars in total).
The IDs come from Lee (1977).

For these targets, Table \ref{Tab6} lists the mean values of the
chemical abundances for CN-S and CN-W(+I) stars. The Na, O, Mg, Al and
Ca abundances are from this work, the $\rm {[C/Fe]}$ values are given
by SB05, while the $\rm {[N/Fe]}$ are the mean values taken from I99 for all
their CN-S and CN-W stars, including stars not in common with our
sample (there are only five stars in our sample with nitrogen content
available). In the last column the differences $\Delta(\rm {S-W})$
between the mean values for CN-S and CN-W(+I) stars are listed.

\begin{table*}[!htpq]
\caption{CN band strengths and Carbon and Nitrogen  
         abundances for our target stars with C, N and CN measurements
         in the literature. We use the 
         star identifications from Lee (1977). In the last column, S
         refers to CN-strong, W CN-weak and I CN-intermediate stars. The C
         and N abundances are from SB05 and I99, respectively.}
\centering
\label{Tab5}
\begin{tabular}{ c l c c c c c l c c c }
\hline\hline
ID  &  S(3839) & CN &${\rm [CI/Fe]}$&${\rm [NI/Fe]}$& &ID   &  S(3839)& CN&${\rm [CI/Fe]}$&${\rm [NI/Fe]}$\\
\hline
L1411&  $0.54 $  & S  &$...  $&$...  $ & &L2626  & $0.66 $ & S&$...  $ &$...  $\\
L1403&  $0.655$  & S  &$-0.67$&$...  $ & &L3705  & $0.20 $ & W&$...  $ &$...  $\\
L1501&  $0.61 $  & S  &$...  $&$...  $ & &L3706  & $0.16 $ & W&$...  $ &$...  $\\
L2422&  $0.69 $  & S  &$-0.83$&$...  $ & &L2519  & $0.39 $ & I&$-0.50$ &$+0.22$\\
L4404&  $0.505$  & S  &$-0.35$&$...  $ & &L2623  & $0.245$ & W&$-0.70$ &$...  $\\
L1514&  $0.305$  & W  &$-1.12$&$+0.99$ & &L3721  & $0.58 $ & S&$...  $ &$ ... $\\
L4415&  $0.53 $  & S  &$-0.67$&$...  $ & &L2617  & $0.63 $ & S&$-0.77$ &$+1.05$\\
L4416&  $0.62 $  & S  &$-0.89$&$...  $ & &L2711  & $0.65 $ & S&$...  $ &$...  $\\
L4509&  $0.62 $  & S  &$-0.44$&$...  $ & &L3730  & $0.24 $ & W&$...  $ &$...  $\\
L1512&  $0.61 $  & S  &$...  $&$...  $ & &L3612  & $0.65 $ & S&$-0.73$ &$+1.02$\\
L1617&  $0.63 $  & S  &$-0.93$&$...  $ & &L3621  & $0.23 $ & W&$...  $ &$...  $\\
L1619&  $0.62 $  & S  &$...  $&$...  $ & &L2608  & $0.57 $ & S&$-0.52$ &$...  $\\
L1608&  $0.60 $  & S  &$-0.63$&$...  $ & &L3624  & $0.40 $ & I&$-0.67$ &$+0.25$\\
L1605&  $0.74 $  & S  &$...  $&$...  $ & &L2410  & $0.23 $ & W&$-0.59$ &$...  $\\
L4630&  $0.505$  & S  &$-0.62$&$...  $ & &L4413  & $0.655$ & S&$-0.32$ &$...  $\\
L3701&  $0.39 $  & I  &$...  $&$...  $ & &L4421  & $0.655$ & S&$-0.83$ &$...  $\\
L3404&  $0.12 $  & W  &$-0.40$&$...  $ & &L4508  & $0.14$ & W&$ ... $ &$...  $\\
\hline
\end{tabular}
\end{table*}

\begin{figure*}
\centering
\includegraphics[width=5cm]{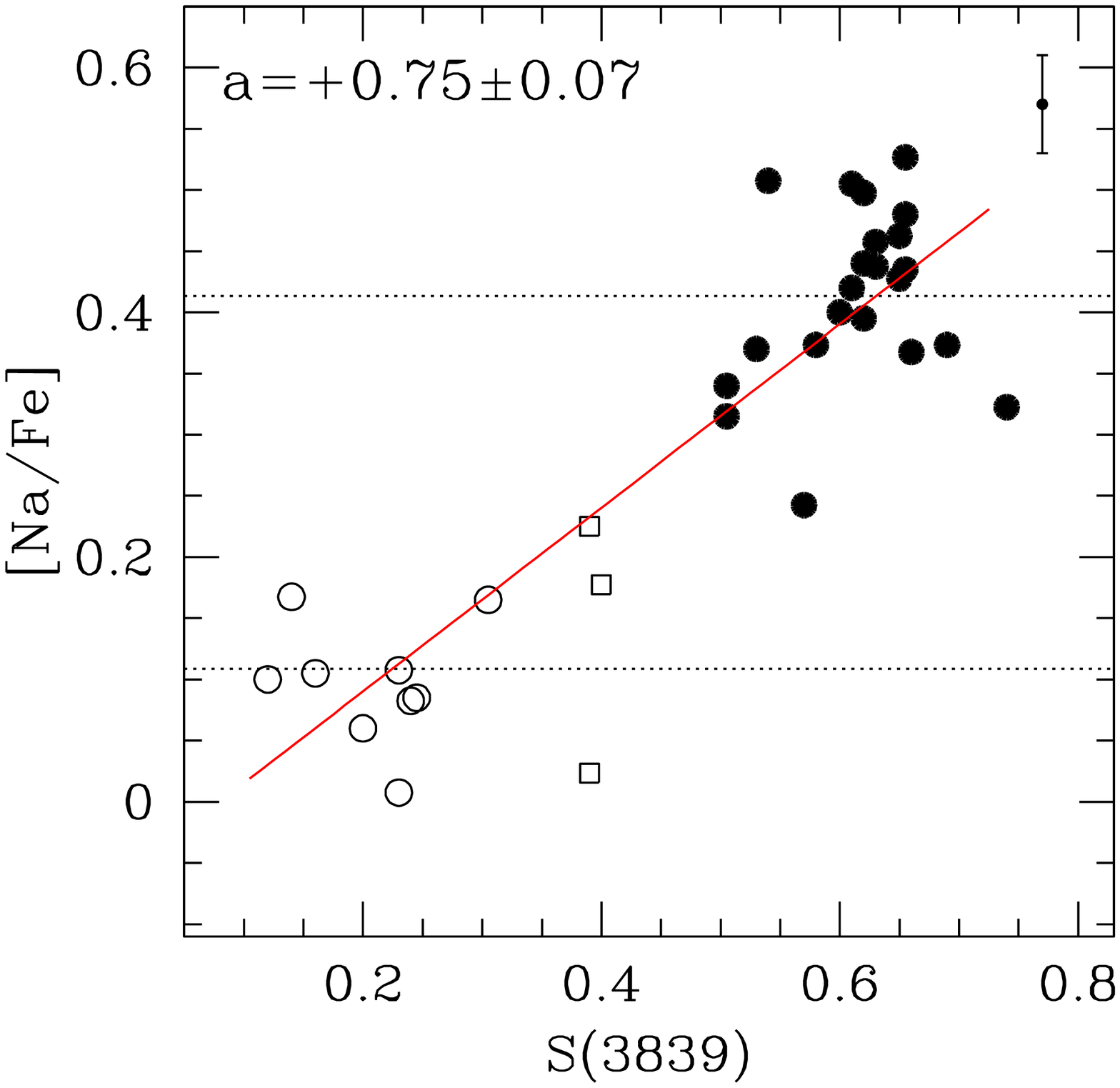}
\includegraphics[width=5cm]{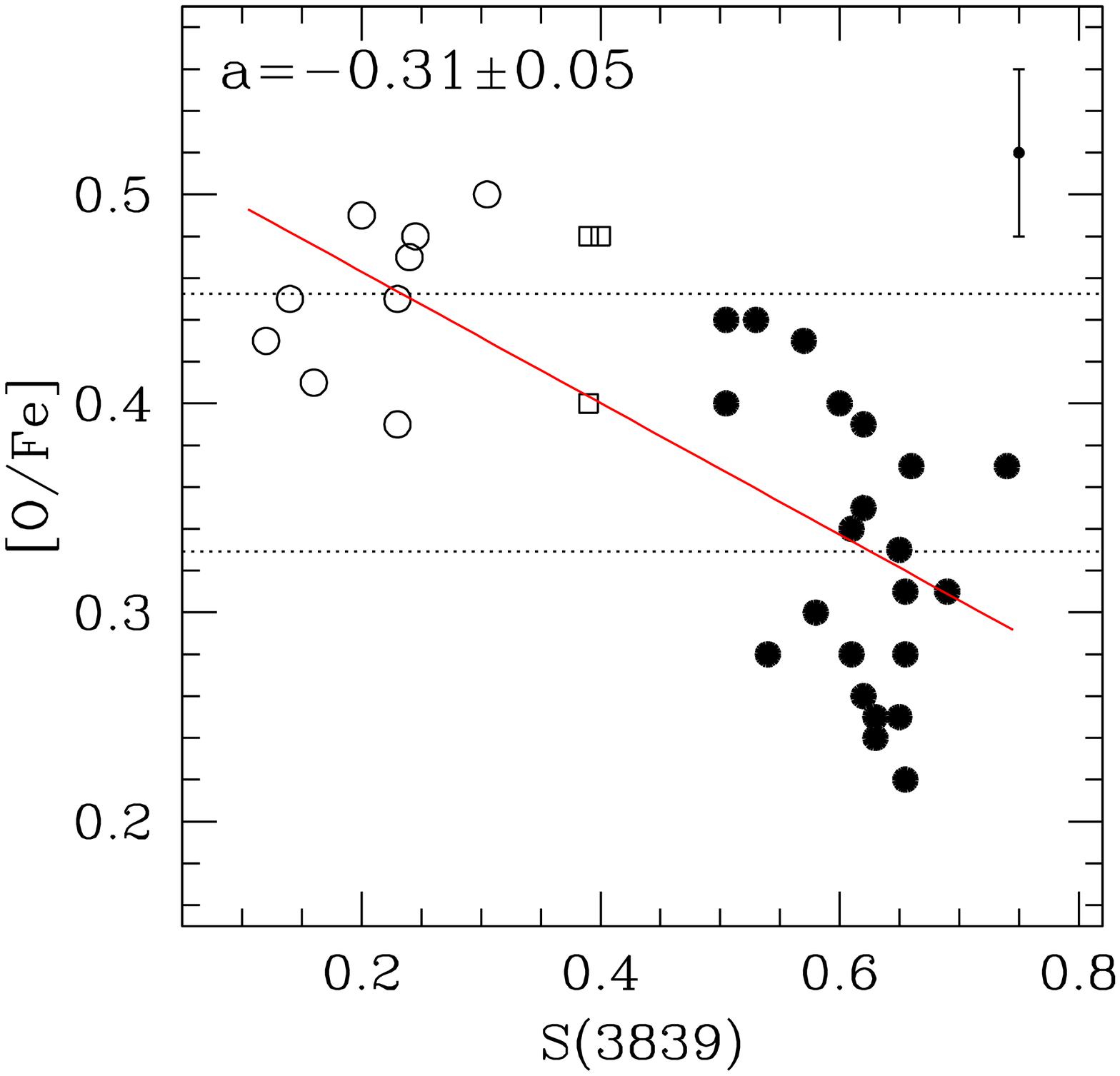}
\includegraphics[width=5cm]{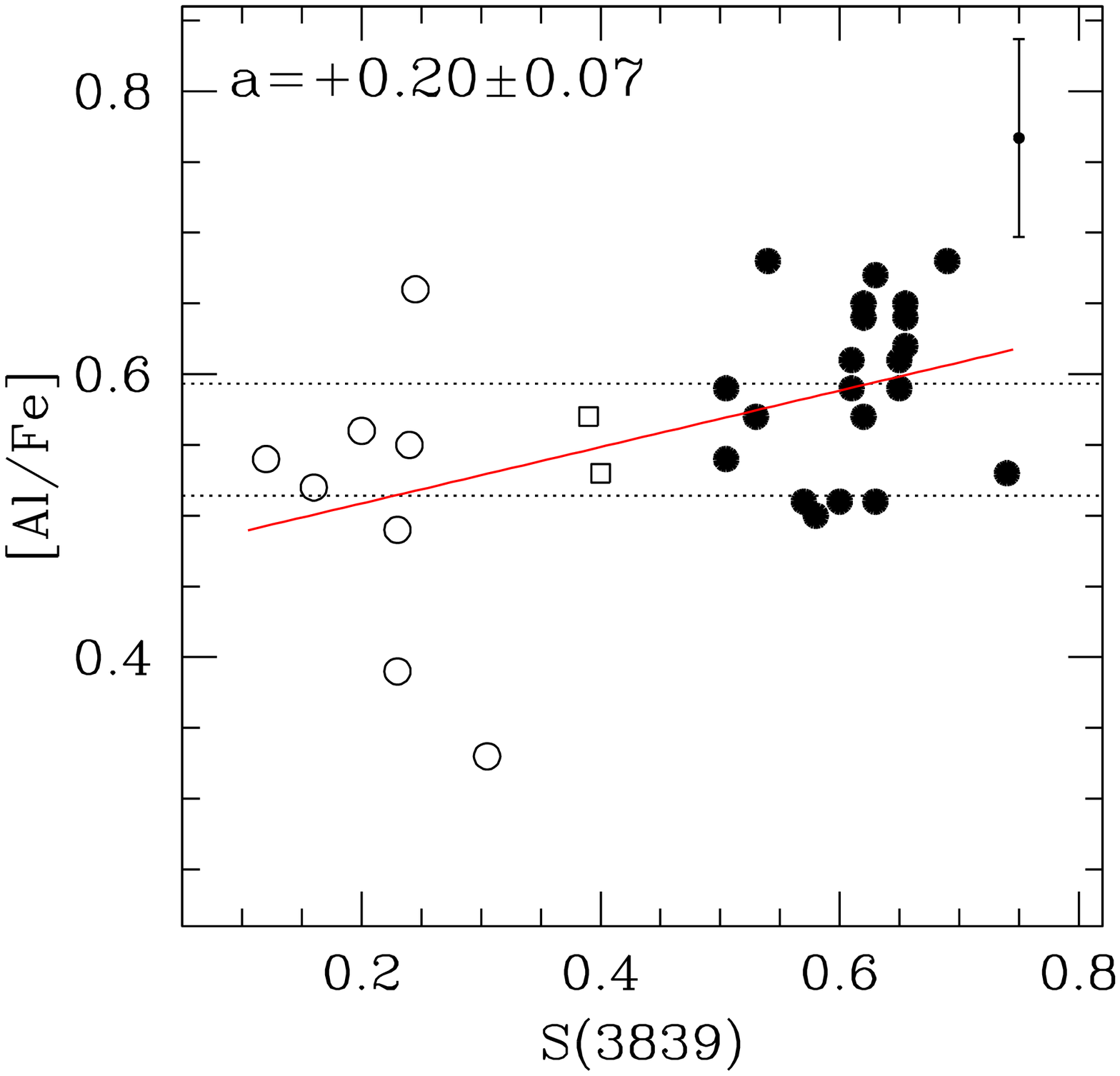}
\includegraphics[width=5cm]{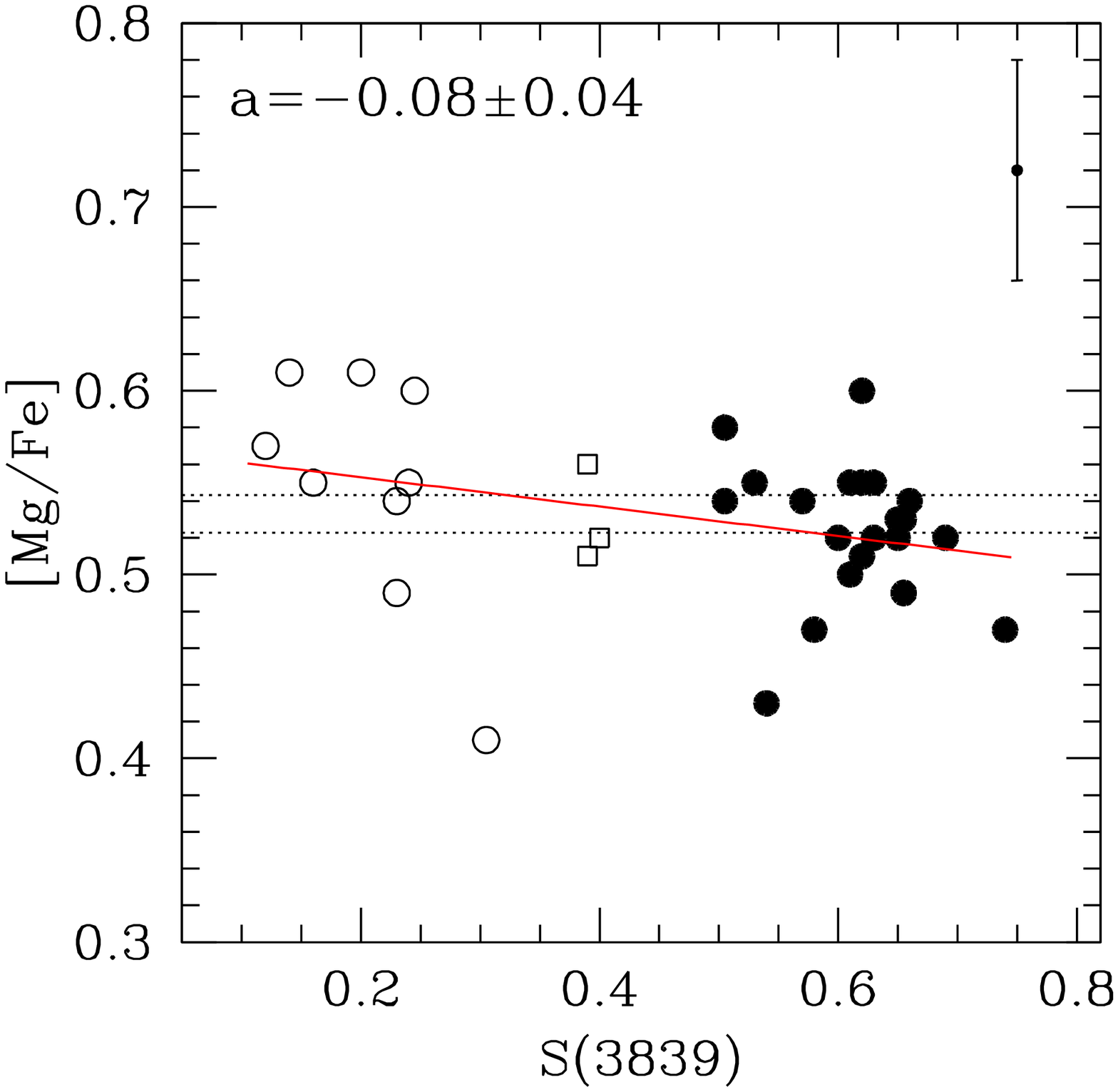}
\includegraphics[width=5cm]{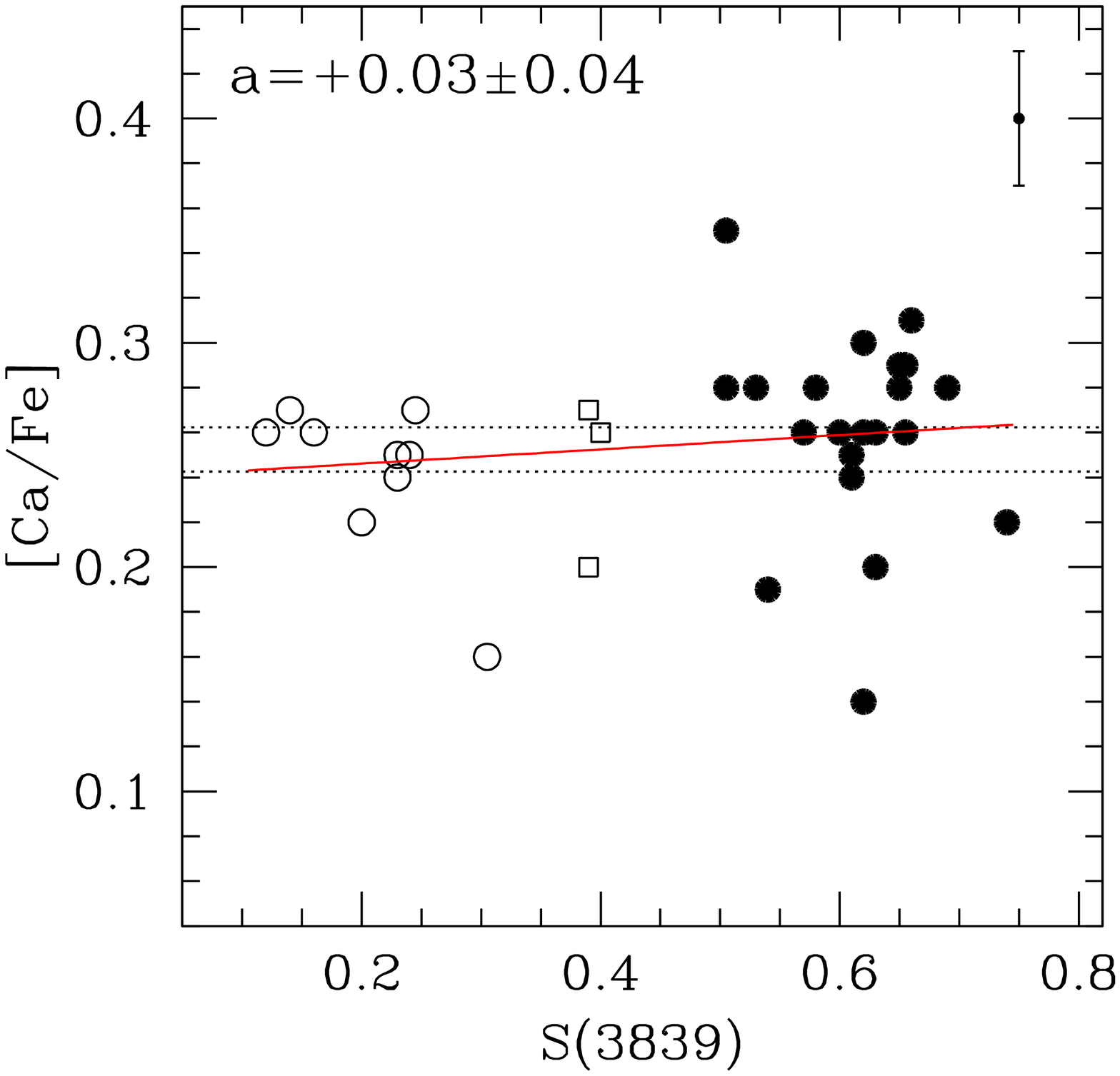}
\caption{The abundance ratios ${\rm [Na/Fe]}$, $\rm {[O/Fe]}$, $\rm
  {[Al/Fe]}$, $\rm {[Mg/Fe]}$ and $\rm {[Ca/Fe]}$ are plotted as a
  function of the CN index S(3839). The filled circles represent the
  CN-S stars, the open circles the CN-W and the open squares the CN-I ones.
  In each panel, $\rm {a}$ is the slope of the straight line of the best
  least square fit. The error bars represent the typical errors
  $\sigma_{\rm tot}$ from Tab.~\ref{Tab3}.}
\label{Fig10}
\end{figure*}

Fig.~\ref{Fig10} shows the ${\rm [Na/Fe]}$, $\rm {[Al/Fe]}$, $\rm
{[O/Fe]}$, $\rm {[Mg/Fe]}$ and $\rm {[Ca/Fe]}$ values
obtained in this work as a function of S(3839). In this
figure CN-S objects are represented by filled circles, CN-W by open
circles, and CN-I by open squares. The dotted lines represent the mean
values of the abundances for the CN-S and CN-W(+I) stars; the CN-I stars
were included in the CN-W group because of their similar behaviour as is
clear from Fig.~\ref{Fig10}. The red
lines represent the best least squares fit of the data, and $\rm {a}$
is the slope of the line.
We note that the CN-S objects clearly show significantly enhanced
Na abundances, while the CN-W(+I) stars have a lower Na content.
The three CN-I stars show low ${\rm [Na/Fe]}$ values typical of CN-weak stars.
There is also a systematic difference in Al between CN-weak and
CN-strong stars, with CN-strong stars richer in Al content. 

\begin{figure*}[!hpbt]
\centering
\includegraphics[width=11.0cm]{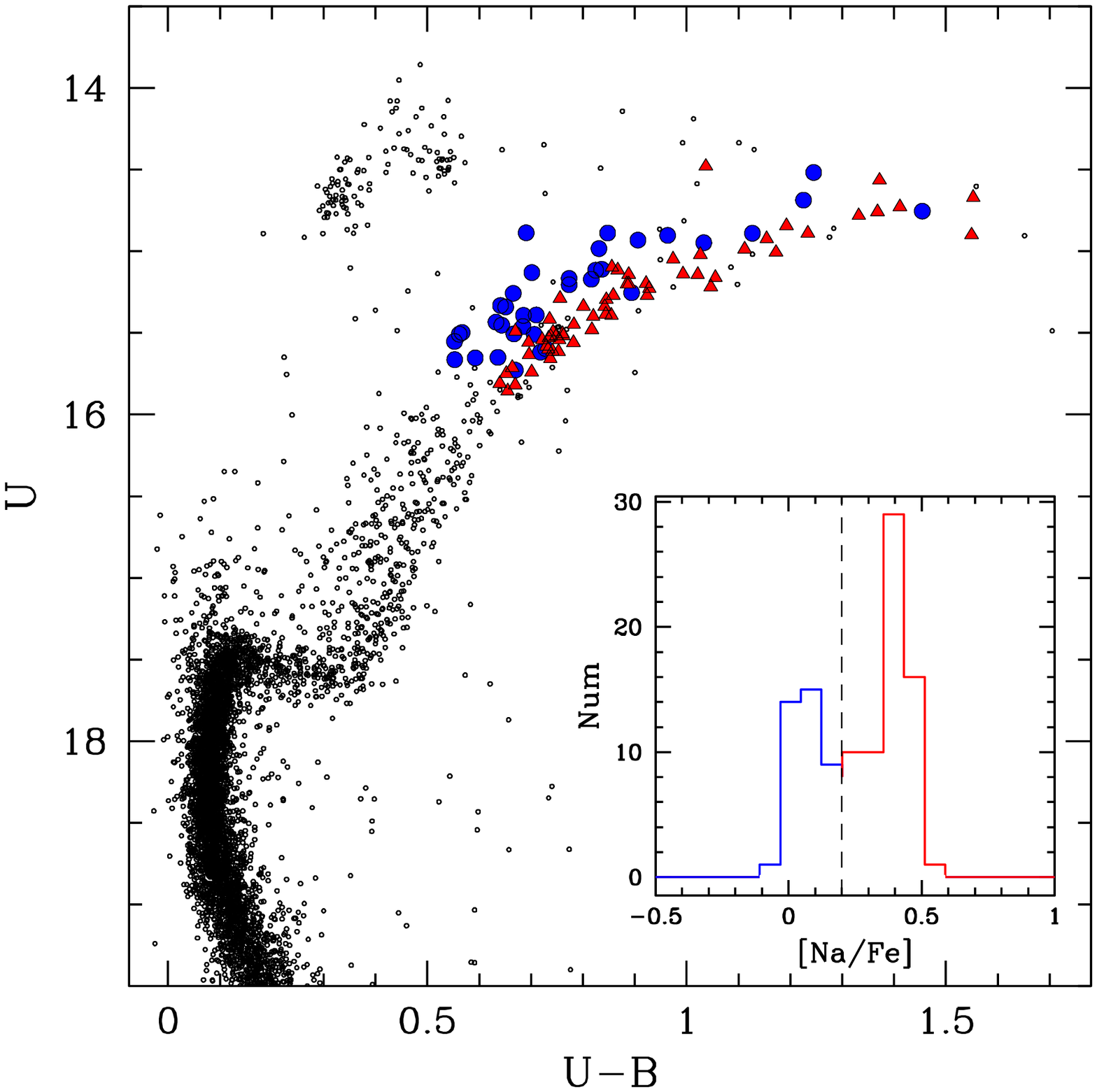}
\caption{$\rm {U}$ vs.\ $\rm {(U-B)}$ CMD from WFI photometry. The distribution of
  the Na content for the UVES stars is shown in the inset. The stars
  belonging to the two different Na groups are represented in two
  different colors: the red triangles represent the stars with $\rm
  {[Na/Fe]}>0.2$ dex, and the blue circles the stars with $\rm
  {[Na/Fe]}\leq0.2$ dex.} 
\label{Fig11}
\end{figure*}

The CN-W(+I) stars have a higher O content than the CN-S as shown in the
upper middle panel of Fig.~\ref{Fig10}.
Fig.~\ref{Fig10} shows some difference in $\rm {[Mg/Fe]}$ 
between the CN-W(+I) and the CN-S stars, with CN-W(+I) stars having
slightly higher Mg abundances than the CN-S ones, but this difference is not
statistically significant (see Table~\ref{Tab6}).

I99 and D92 also found that the scatter in Ca abundance
correlates with the CN strength. I99 found a difference in Ca content
between CN-strong and CN-weak stars by $0.08\pm0.11$ dex.
To compare, we have also analysed the Ca abundance as a function of the CN
strength.
In the bottom-right panel of Fig.~\ref{Fig10}, $\rm {[Ca/Fe]}$ is plotted as a
function of S(3839); we can see that the difference between the two
groups is 0.02 dex, smaller than what was found by I99 
and not statistically significant (see Table~\ref{Tab6}). 

\begin{table*}[!htpq]
\caption{Mean abundance ratios and relative rms ($\sigma$) for the
  CN-S and CN-W(+I) groups. In the last column the difference between
  CN-S and CN-W(+I) stars are listed.}
\centering
\label{Tab6}
\begin{tabular}{lccccccc}
\hline\hline
 [el/Fe]        & CN-S       &$\sigma$&    &CN-W(+I) &$\sigma$&&$\Delta({\rm S-W(+I)})$\\
\hline
${\rm [CI/Fe]}$  &$-0.66\pm 0.05$ &0.19&    &$-0.66 \pm 0.11$ &0.25&& $+0.00$  \\
${\rm [NI/Fe]}$  &$+1.08\pm 0.10$ &0.22&    &$+0.42 \pm 0.19$ &0.42&& $+0.66$  \\
${\rm [OI/Fe]}$  &$+0.33\pm 0.01$ &0.07&    &$+0.45 \pm 0.01$ &0.04&& $-0.12$  \\
${\rm [NaI/Fe]}$ &$+0.41\pm 0.02$ &0.07&    &$+0.11 \pm 0.02$ &0.06&& $+0.30$  \\
${\rm [AlI/Fe]}$ &$+0.59\pm 0.01$ &0.06&    &$+0.51 \pm 0.03$ &0.09&& $+0.08$  \\
${\rm [MgI/Fe]}$ &$+0.52\pm 0.01$ &0.04&    &$+0.54 \pm 0.02$ &0.06&& $-0.02$  \\
${\rm [CaI/Fe]}$ &$+0.26\pm 0.01$ &0.04&    &$+0.24 \pm 0.01$ &0.03&& $+0.02$  \\
\hline
\end{tabular}
\end{table*}

\section{Search for evolutionary effects}

\subsection{Abundances along the RGB}

\begin{figure}[!hpbt]
\centering
\includegraphics[width=8.5cm]{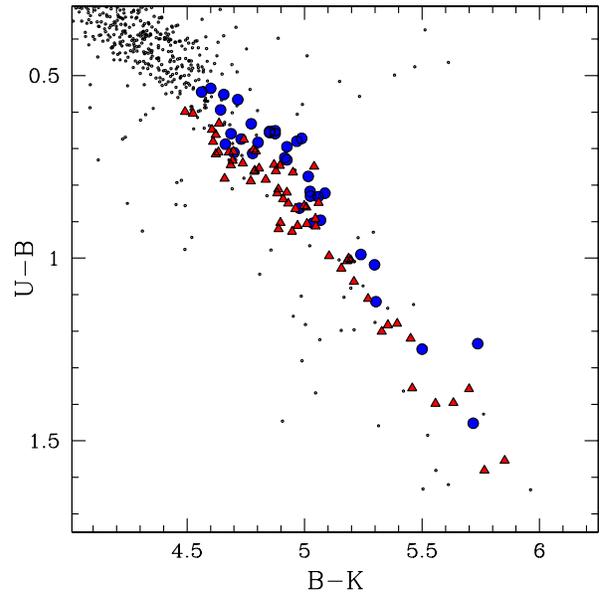}
\caption{$\rm {(U-B)}$ vs.\ $\rm {(B-K)}$ diagram: $\rm {U}$ and $\rm {B}$ come
  from WFI photometry, $\rm {K}$ from 2MASS catalogue. The UVES stars
  belonging to the two different Na groups are represented as in
  Fig.~\ref{Fig11}. }
\label{Fig12}
\end{figure}

According to the results discussed in the previous section, we can
define two groups of stars: the Na-rich stars, i. e. those with $\rm
{[Na/Fe]}\geq0.2$ and low Oxygen content which must be associated
to the CN-S stars (Fig.~\ref{Fig10}), and the Na-poor stars with
higher Oxygen, i. e. those with $\rm {[Na/Fe]<0.2}$, which correspond
to the CN-W group. It is instructive to look at the position of the
stars belonging to these two Na groups on the $\rm {U}$ vs.\ $\rm
{(U-B)}$ CMD and on the color-color $\rm {(U-B)}$ vs.\ $\rm {(B-K)}$ 
diagram (Fig.~\ref{Fig11} and Fig.~\ref{Fig12}, respectively). The
$\rm {U}$ and $\rm {B}$ magnitudes come from our WFI photometry, the
$\rm {K}$ magnitude from the 2MASS catalogue.  
The Na-rich stars are represented by red triangles, while the Na-poor
stars by blue circles. The CMD has been corrected for differential reddening
(following the same procedure used in Sarajedini et al.\ 2007) and
proper motions. 
It is clear from Fig.~\ref{Fig11} that our sample of stars mostly
includes RGB stars with only three or four probable AGB stars.

Interestingly enough, the two groups of Na-rich and Na-poor stars form
two distinct branches on the RGB.  Na-rich stars define a narrow
sequence on the red side of the RGB, while the Na-poor sample
populates the blue, more spread out portion of the RGB.  Even more
interestingly, the anomalous broadening of the RGB is visible down to
the base of the RGB, at $\rm {U} \sim 17.5$, indicating that the two
abundance groups are present all over the RGB, even well below the RGB
bump, where no deep mixing is expected. This evidence further
strengthens the idea that the bimodal Na, O, CN distribution must have
been present in the material from which the stars we presently observe
in M4 originated.

\begin{figure}[hpbt]
\centering
\includegraphics[width=8.5cm]{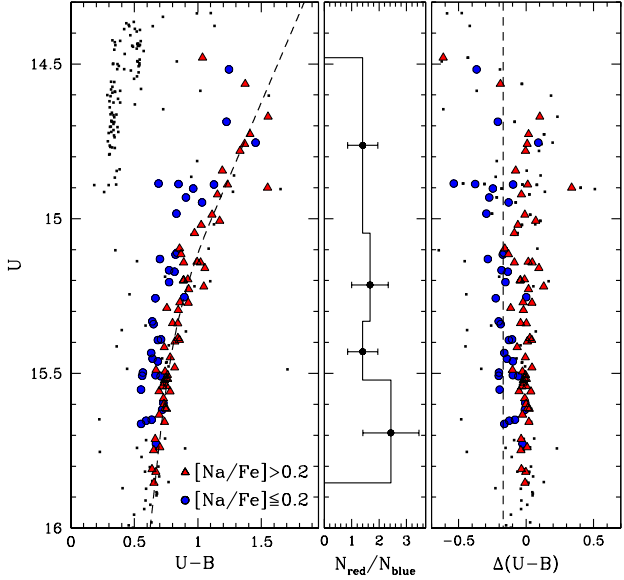}
\caption{(Left panel) Distribution along the RGB of the stars with measured
  abundances. The Na-rich (red) and the Na-poor (blue) stars are represented as in
  Fig.~\ref{Fig11} and Fig.~\ref{Fig12}. In the middle panel the $\rm
  N_{red}/N_{blue}$ ratios for the 4 selected magnitude bins are shown
  with their relative Poisson errors. The rightmost panel shows the color
  difference between each analyzed star and a reference fiducial line (dotted
  line).}
\label{Fig13}
\end{figure}

Figure~\ref{Fig11} unequivocally shows that the Na (CN) dichotomy
is associated to a dichotomy in the color of the RGB stars.
In order to quantify this split, we calculated for the Na-poor group of stars
the mean difference in $\rm {(U-B)}$ color with respect to a fiducial
line representative of the RGB Na-rich stars (see Fig.~\ref{Fig13}). 
The fiducial line has been obtained as follows:

\begin{itemize}

\item{the CMD has been divided into bins of 0.30 mag in $\rm {U}$, and a median
  color $\rm {(U-B)}$ and $\rm {U}$ magnitude have been computed for the
  Na-rich ($\rm {[Na/Fe]}\geq0.2$) stars in each bin. The median points have been fitted
  with a spline function which represent a first, raw fiducial line;}

\item{for each Na-rich star, the difference in color with respect to the
  fiducial was calculated, and the 68.27th percentile of the absolute
  values of the color differences, was taken as an estimate of the
  color dispersion ($\sigma$). All stars with a color distance from the fiducial
  larger than $3\sigma$ were rejected;}

\item{the median colors and magnitudes and the $\sigma$ of each bin
were redetermined by using the remaining stars.}

\end{itemize}

The leftmost panel of Fig.~\ref{Fig13} shows the original CMD, while
the rightmost one shows the CMD after subtracting from each Na-poor star
the fiducial line color appropriate for its ${\rm U}$ magnitude. The
differences are indicated as $\rm {\Delta(U-B)}$.
We calculated the average $\rm {\Delta(U-B)}$ of Na-poor
stars with $\rm {U}$$\geq$14.8. This cut in magnitude has been imposed
by the poor statistics and the
presence of the probable AGB stars at brighter magnitudes. The
mean $\rm {\Delta(U-B)}$ value is $-0.17\pm0.02$ (dotted line in the
left most panel of Fig.~\ref{Fig13}).

The significant difference between the mean colors of the 2 groups of
stars is a further evidence of the presence of two different 
stellar populations in the RGB of M4, to be associated
with the different content of Na, and, because of the discussed
correlations to different O, N, and C content.

\begin{figure*}[!hpbt]
\centering
\includegraphics[bb=0 220 550 600,width=8.6cm]{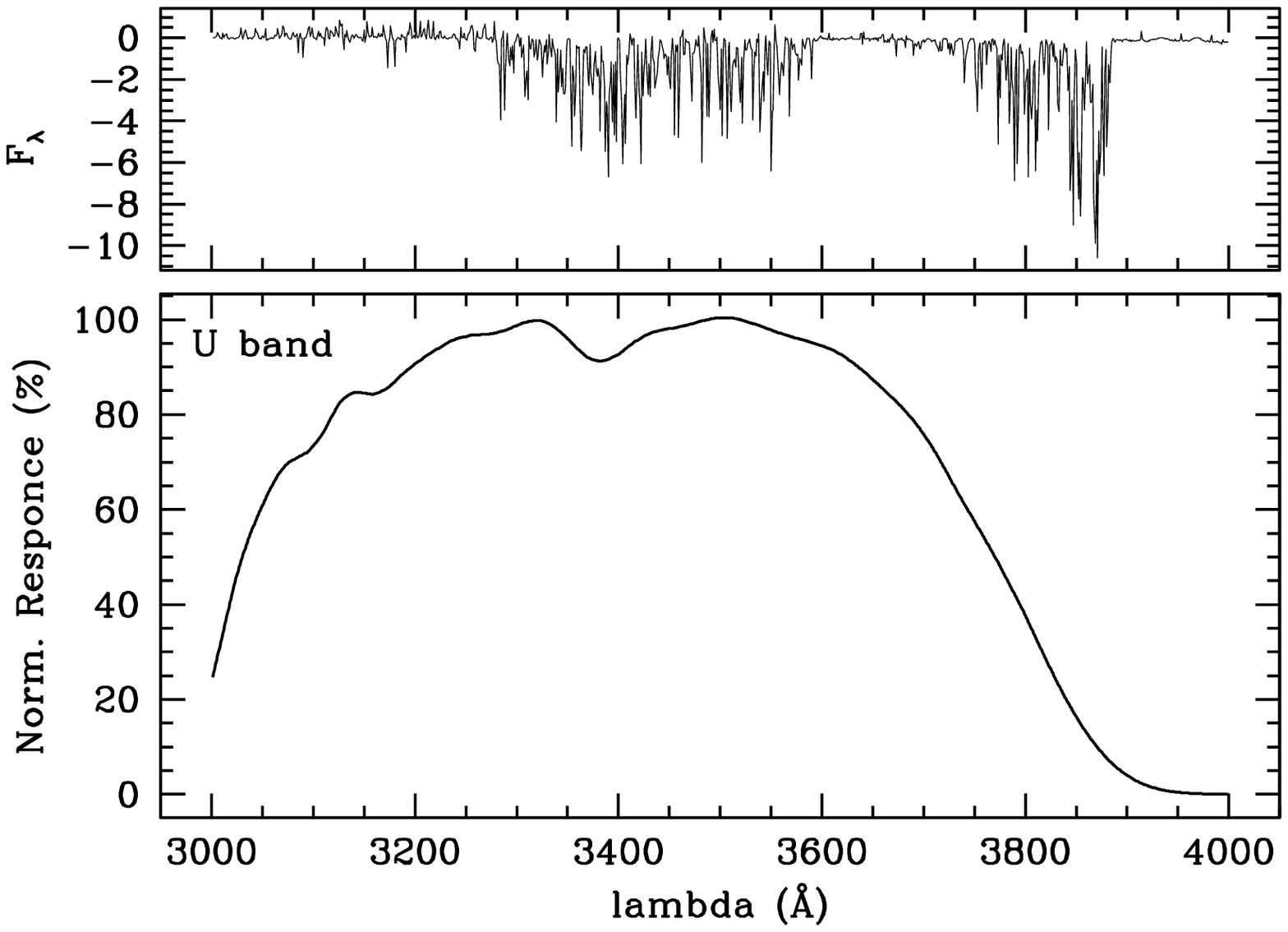}
\includegraphics[bb=0 220 550 600,width=8.6cm]{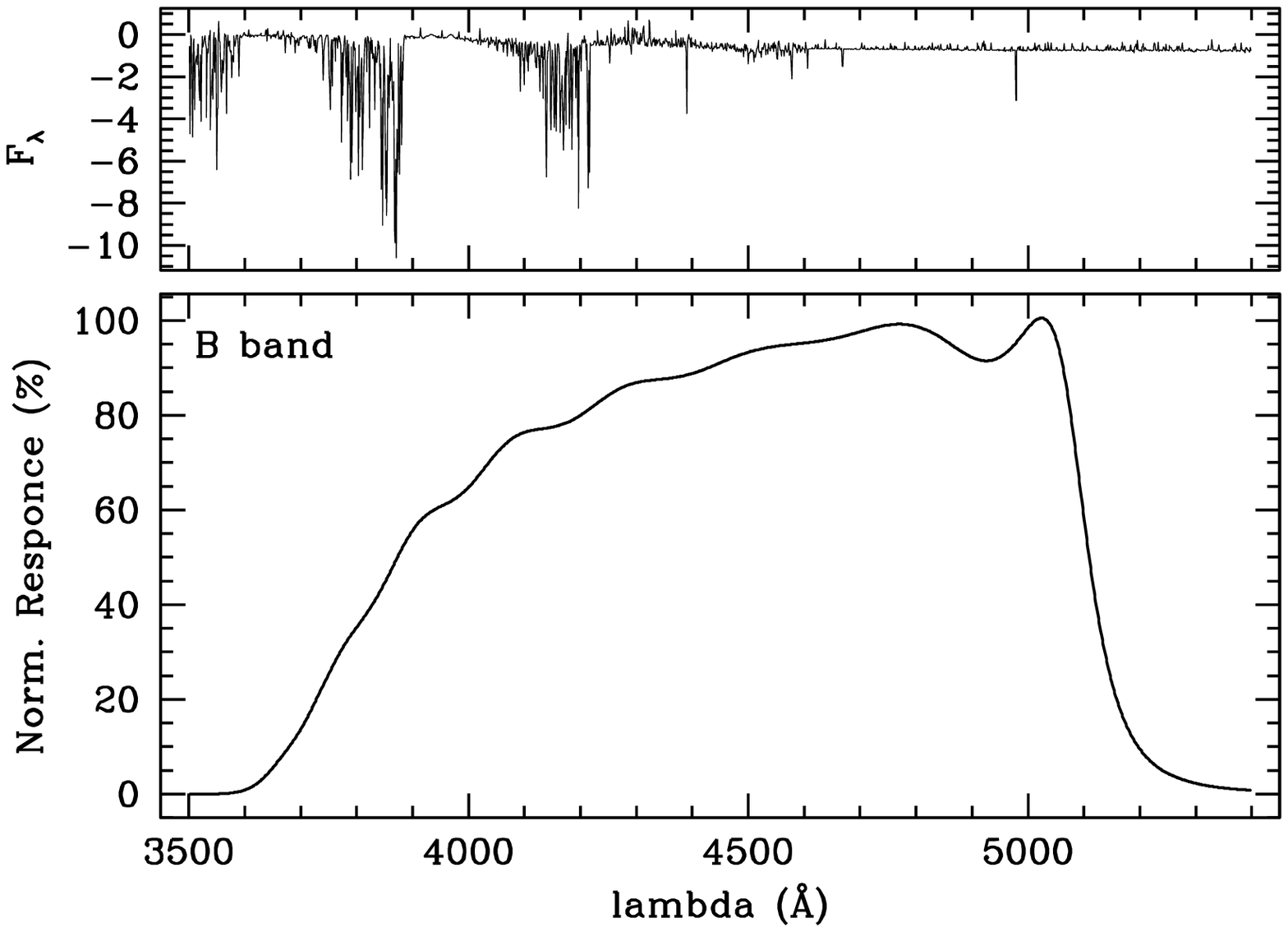}
\caption{The differences (expressed in $\rm
  {10^{4}\ erg/cm^2/sec/Angstrom}$) between the CN-S and the CN-W(+I)
  simulated spectra in the $\rm {U}$ and $\rm {B}$ band are represented in the
  upper panels. The normalized response of the two filters is also
  shown in the lower panels.}  
\label{Fig14}
\end{figure*}

In order to understand the origin of the photometric dichotomy, using
SPECTRUM, we simulated two synthetic spectra, one representative of
the Na-rich, CN-strong stars and one for the Na-poor, CN-weak ones.
The two synthetic spectra were computed using as atmospheric parameters the mean
values measured for the sample of our stars for which there are literature
data on the CN band strengths (see Tab.~\ref{Tab5}), and assuming for the C, N,
O, Na abundances the average values calculated for the CN-W(+I) and
the CN-S groups and listed in Tab.~\ref{Tab6}. We then multiplied the two
spectra by the efficency curve of our $\rm {U}$ and $\rm {B}$ photometric bands
(Fig.~\ref{Fig14}, lower panels), and, finally, we calculated the
difference between the resulting fluxes.
The upper panels of Fig.~\ref{Fig14} show the differences between the
two simulated spectra as a function of the wavelength for the $\rm
{U}$ (left panel) and $\rm {B}$ (right panel) panel.
It is clear that the strength of the CN and NH bands strongly influences the
$\rm {U} - {B}$ color. The NH
band around 3360 \AA, and the CN bands around 3590 \AA, 3883 \AA\ and
4215 \AA \ are the main contributors to the effect.

The differences in magnitude between the CN-S and the CN-W(+I)
simulated spectra in the two bands are: $\rm
{\Delta(U)_{Strong-Weak}=+0.06}$ and $\rm
{{\Delta(B)_{Strong-Weak}=+0.02}}$. Consequently, the expected color
difference between the two groups of stars is $\rm
{{\Delta(U-B)_{Strong-Weak}=+0.04}}$. 
This value goes in the same direction, but it is smaller
than the observed one ($\rm {\Delta(U-B)_{Strong-Weak}=+0.17\pm0.02}$).
However, we note that our procedure uses simulated
spectra with average abundances, and therefore should
be considered a rough simulation.
We cannot exclude the possibility that other effects (perhaps related to the
structure and evolution of the stars with different chemical content, or
effects on the stellar atmosphere associated with the complex distribution of
the chemical abundances in addition to the CN band), might further contribute
to the photometric dichotomy in the RGB.
Surely, our simulations show that the CN-bimodality affects the
$\rm {(U - B)}$ color and can be at least partly responsible for the
observed spread in the $\rm U$ vs.\ $\rm {(U - B)}$ CMD.

Finally, we investigated whether the chemical and photometric dichotomy
is related to the evolutionary status along the RGB.
To this end, we have calculated the fractions of stars in the 2
Na-groups at different magnitudes along the RGB.
As shown in the middle panel of Fig.~\ref{Fig13},
we divided the RGB in 4 magnitude bins
containing the same number of stars with measured metal content,
and calculated the ratio between the number of Na-rich ($\rm
{N_{red}}$) and the number of Na-poor ($\rm {N_{blue}}$) stars in each
bin. The $\rm {N_{red}/N_{blue}}$ ratios with their associated Poisson errors
as a function of the magnitude are plotted the middle panel of Fig.~\ref{Fig13}.
The $\rm {N_{red}/N_{blue}}$ values in the different magnitude
bins are the same within the errors.
A similar result is obtained if we divide the RGB in two bins only,
but containing a number of
stars twice as large as in the previous experiment.
In this case, we have $\rm {N_{red}/N_{blue}} = 1.52\pm0.44$ for $\rm
{14.0<U<14.9}$, $\rm {N_{red}/N_{blue}} = 1.82\pm0.54$ for $\rm
{14.9<U<15.8}$.

\begin{figure*}[!hpbt]
\centering
\includegraphics[bb=0 250 550 620,width=13.5cm]{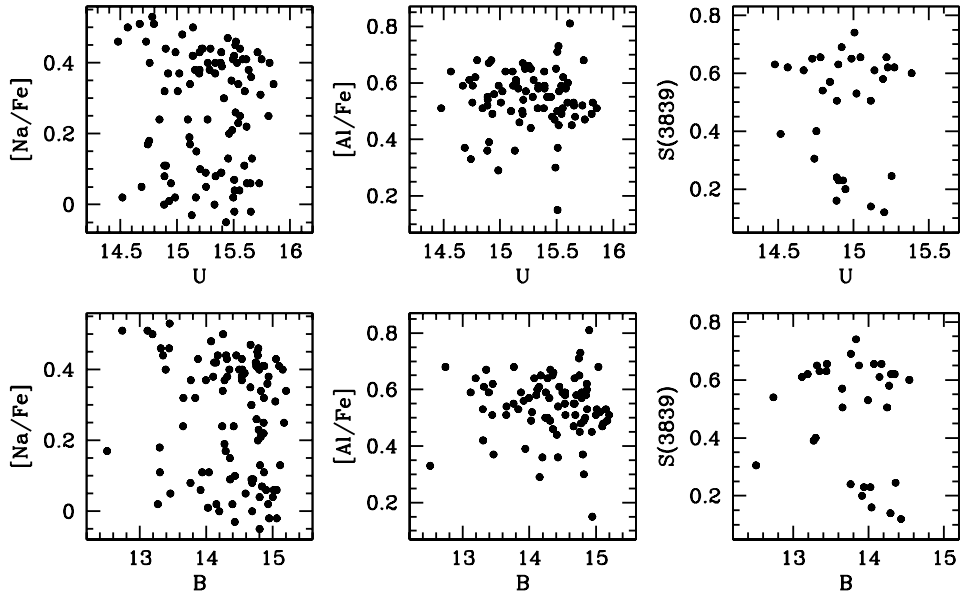}
\caption{Dependence of Na, Al, and S(3839) index as
 a function of the magnitude $\rm {U}$ (upper panels) and $\rm {B}$
 (lower panels).} 
\label{Fig15}
\end{figure*}

Figure~\ref{Fig15} shows the dependence of Na, Al, and S(3839) index as
a function of the magnitude $\rm {U}$ (upper panel) and $\rm {B}$
(lower panel). Again, we do not see any trend of the
abundances with the position of the stars along the RGB.

In conclusion, there is no evidence for a dependence of the Na content
and (because of the discussed correlations) of the O, Al abundances
or of the CN strength with the evolutionary status of the stars.

\subsection{The RGB progeny on the HB}

In the previous sections, we have identified two groups of stars
with distinct Na, O, CN content, and which populate two
distinct regions in the $\rm {U}$ vs.\ $\rm {U-B}$ CMD.

We note that Yong \& Grundhal (2008), by studying 8 bright red giants
in the GC NGC~1851 found a large star to star abundance variation in
the Na, O, and Al content, and suggested that these abundance
anomalies could be associated with the presence of the two stellar
populations identified by Milone et al. (2008), a situation apparently
very similar to what we have found in M4. Cassisi et al. (2008)
suggest that extreme CNO variations could be the basis of the SGB
split found in NGC~1851. As discussed in Milone et al. (2008) there is
some evidence of a split of the RGB in NGC~1851 similar to what we
have found in this paper for M4.  
We have investigated all available HST data of M4, including
CMDs from very high precision photometry based on ACS/HST images; we could 
find no evidence for a SGB split (as in NGC~1851)
or evidence for a MS split (as in NGC~2808 or $\omega$~Cen).

It is also useful to investigate where the progeny of the two RGB
populations is along the HB.
Milone et al.\ (2008) suggest that the bimodal HB of NGC~1851 can be
interpreted in terms of the presence of two distinct stellar
populations in this cluster,
and tentatively link the bright SGB to the red HB (RHB) and the fainter
SGB to the blue HB  (BHB). M4 also has a bimodal
HB, well populated on both the red and blue side of the RR Lyrae gap.
Could the HB morphology of M4 be related to the CN bimodality? N81,
studying the CN band strengths on a sample of 45 giant stars in M4,
suggested the possibility of a relation between the HB morphology and the
CN-bimodality.

To test the possible relation between the chemical and consequent RGB
bimodality of M4 and the morphology of its HB, we used the WFI@2.2m
instrumental photometry by Anderson et al.\ 2006 (Fig.~\ref{Fig16})
corrected for differential reddening, and selected the cluster members using the
measured proper motions.  
We remark here the fact that this photometry is not the one we use in the
whole paper. The homogenization between the two photometries is
a difficult and long process (mainly because the fact that the
$\rm {B}$ filters for the two datasets are different) beyond the scope of this work.  

In our test, we used the $\rm {V}$ vs.\ $\rm {(B-V)}$ CMD, as in this
photometric plane the two components of the HB are more clearly
distinguishable.

In Fig.~\ref{Fig16}, different symbols show the red and blue HB stars
that we selected. The
ratio between the number ({$\rm N_{BHB}$}) of the HB stars bluer than
the RR-Lyrae instability strip (BHB), and the total HB stars ({$\rm
  N_{TOT}$) is: 

\begin{center}
${\rm N_{BHB}/N_{TOT} = 0.56 \pm 0.10}$
\end{center}

\noindent
while the ratio between the Na rich stars (${\rm N_{NaR}}$) and the
total number of stars in our spectroscopic sample (${\rm N_{S}}$) is:

\begin{center}
${\rm N_{NaR}/N_{S} = 0.64 \pm 0.10}$
\end{center}

where the associated uncertainties are the Poisson errors.
Within the statistical uncertainties, accounting for the different
evolutionary times along the HB,
we can tentatively associate the Na-rich stars with the  BHB, and the
Na-poor stars with the RHB. 
A direct measurement of the metal content of the HB stars is needed in
order to support this suggestion..

\begin{figure}[hpbt]
\centering
\includegraphics[width=9cm]{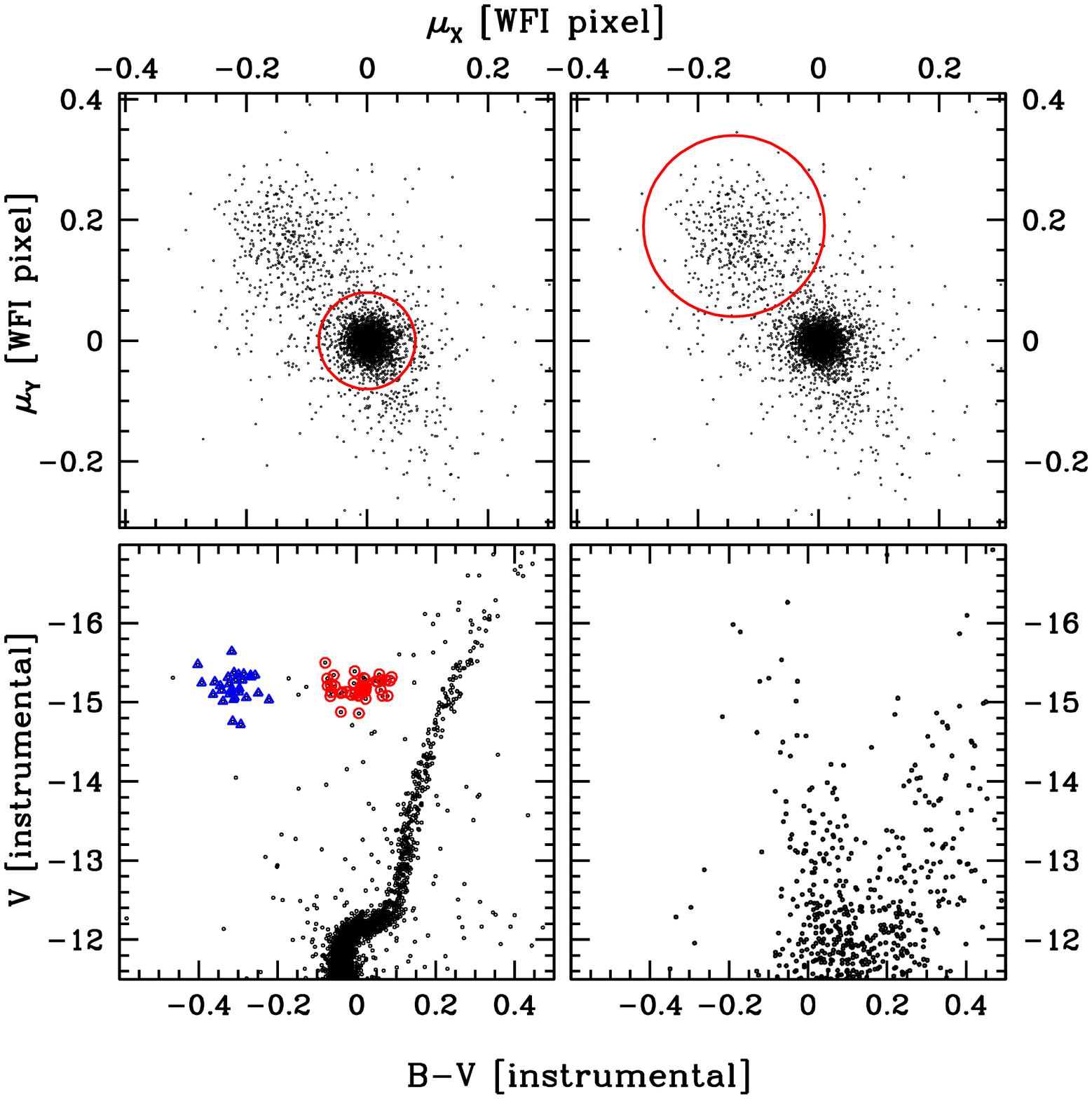}
\caption{In the top panels, the proper motions diagrams show the
separation of the probable cluster stars (left panel) from the field
(right panel). In the bottom panels, the cluster and field CMD ($\rm
{V}$ vs.\ $\rm {B-V}$) are shown for the stars within the red circle in
their respective upper panels. In the M4 CMD, the stars on the red and
on the blue side of the RR Lyrae instability strip are represented
with red and blue symbols respectively. For this figure, the photometry, which
is not photometrically calibrated, is taken from Anderson et al.\ (2006).}
\label{Fig16}
\end{figure}

\begin{table*}[!htpq]
\caption{Mean abundance ratios and relative rms ($\sigma$) for the
  Na-rich and Na-poor groups. In the last column, the difference
  between Na-rich and Na-poor stars are listed.} 
\centering
\label{Tab7}
\begin{tabular}{lccccccc}
\hline\hline
 [el/Fe]        & Na-rich        &$\sigma$&    &Na-poor          &$\sigma$&&$\Delta({\rm Na_{rich-poor}})$\\
\hline
${\rm [OI/Fe]}$     &$+0.34\pm 0.01$ &0.08  &    &$+0.47\pm 0.01$ &0.04&&$-0.13\pm0.01$  \\
${\rm [NaI/Fe]}$    &$+0.38\pm 0.01$ &0.08  &    &$+0.07\pm 0.01$ &0.06&&$+0.31\pm0.01$  \\
${\rm [MgI/Fe]}$    &$+0.48\pm 0.01$ &0.06  &    &$+0.52\pm 0.01$ &0.05&&$-0.04\pm0.01$  \\
${\rm [AlI/Fe]}$    &$+0.57\pm 0.01$ &0.08  &    &$+0.49\pm 0.02$ &0.13&&$+0.08\pm0.02$  \\
${\rm [SiI/Fe]}$    &$+0.49\pm 0.01$ &0.04  &    &$+0.46\pm 0.01$ &0.06&&$+0.03\pm0.01$  \\
${\rm [CaI/Fe]}$    &$+0.28\pm 0.01$ &0.04  &    &$+0.28\pm 0.01$ &0.04&&$+0.00\pm0.01$  \\
${\rm [TiI/Fe]}$    &$+0.29\pm 0.01$ &0.03  &    &$+0.29\pm 0.01$ &0.05&&$+0.00\pm0.01$  \\
${\rm [TiII/Fe]}$   &$+0.35\pm 0.01$ &0.07  &    &$+0.34\pm 0.01$ &0.05&&$+0.01\pm0.01$ \\
${\rm [CrI/Fe]}$    &$-0.04\pm 0.01$ &0.05  &    &$-0.03\pm 0.01$ &0.05&&$-0.01\pm0.01$  \\
${\rm [FeI/H]}$     &$-1.07\pm 0.01$ &0.05  &    &$-1.07\pm 0.01$ &0.05&&$+0.00\pm0.01$  \\
${\rm [NiI/Fe]}$    &$+0.02\pm 0.01$ &0.03  &    &$+0.01\pm 0.01$ &0.03&&$+0.01\pm0.01$  \\
${\rm [BaII/Fe]}$   &$+0.42\pm 0.01$ &0.07  &    &$+0.40\pm 0.02$ &0.13&&$+0.02\pm0.02$ \\
\hline
\end{tabular}
\end{table*}

\section{Conclusions}

We have presented high resolution spectroscopic analysis of 105 RGB
stars in the GC M4 from UVES data.

We have found that M4 has an iron content $\rm {[Fe/H]=-1.07\pm0.01}$
(the associated error here is the internal error only),
and an $\alpha$ element overabundance $\rm
{[\alpha/Fe]=+0.39\pm0.05}$.
Si and Mg are more overabundant than the other $\alpha$
elements, suggesting a primordial overabundance for these elements.
Moreover, we also find a slight overabundance of Al.
The ${\rm [Na/Fe]}$ versus [O/Fe] ratios follow the well-known Na-O
anticorrelation, signature of proton capture reactions at high
temperature. No Mg-Al anticorrelation was found.

We find a strong dichotomy in Na abundance, and show that it must be
associated to a CN bimodality. Tab.~\ref{Tab7} synthesizes the
spectroscopic and photometric differences we found for the two groups
of stars. Comparing our results to those in the literature on
the CN band strength, the CN-strong stars appear to have higher
content in Na and Al, and lower O than the CN-weak ones. Apparently,
M4 hosts two different stellar populations. This fact is evident from
a chemical abundance distribution, and is also confirmed by
photometry. In fact, an inspection of the $\rm {U}$ vs.\ $\rm {(U-B)}$
CMD reveals a broadened RGB, and, as shown in our investigation, the
two Na groups of stars occupy different positions (have different
colors) along the RGB. Our sample of stars is composed of objects with
$\rm {U}$ brighter than $\sim$16, but, as shown in Fig.~\ref{Fig11},
the RGB appears to be similarly spread from the level of the SGB to the RGB
tip. Since our photometry has been corrected for differential
reddening, we conclude that the observed RGB spread at fainter magnitudes must
also be due to the metal content dichotomy.

We did not find any evident dependence of the chemical abundance
distribution on the evolutionary status (along the RGB) of the target
stars, from below the RGB bump to the RGB tip. This is an additional
indication that the abundance spread must be primordial.

The two groups of stars we have identified both spectroscopically
and photometrically seem to be due to the presence of two distinct
populations of stars. 
The abundance anomalies are very likely due to primordial variations in the
chemical content of the material from which M4 stars formed, and not to
different evolutionary paths of the present stellar population of M4.
This is somehow surprising, because of the relatively small mass of M4 -- it
is an order of magnitude smaller than the mass of $\omega$~Cen,
NGC~2808, NGC~1851, NGC~6388, the other clusters in which a multiple
stellar generation has been confirmed. 
Where did the gas which polluted the material for the CN-Na rich
stars come from? Has it been ejected from a first generation of
stars? How could it stay within the shallow gravitational potential of M4?

The idea that M4 hosts two generations of stars makes the multipopulation
phenomenon in GCs even more puzzling than originally thought. It
becomes harder and harder to accept the idea that the phenomenon can
be totally internal to the cluster, unless this object is what remains
of a much larger system (a larger GC or the nucleus of a dwarf galaxy?).
Surely, Because its orbit involves frequent passages at high inclination
through the Galactic disk, always at distances from the Galactic center
smaller than 5 kpc (see Dinescu et al.\ 1999), M4 must have been
strongly affected by tidal shocks, and therefore it might have been
much more massive in the past.

%\begin{acknowledgements}
%
%\end{acknowledgements}

%\end{document}

\longtab{8}{
\begin{longtable}{lcccccccccccc}
\caption{\label{Tab8} Coordinates, atmospheric parameters, U,B,V,I$_{\rm C}$,\
  and 2MASS J,H,K magnitudes for the analised stars. All magnitudes are not corrected for differential reddening.}\\
\hline
\hline
ID & RA($^0$) & DEC($^0$) & T$_{\rm eff}$ (K)& log(g) &  v$_{\rm t}$ (km/s) & U & B & V & I$_{\rm C}$ & J & H & K \\
\hline
\endfirsthead
\caption{continued.}\\
\hline\hline
ID & RA($^0$) & DEC($^0$) & T$_{\rm eff}$ (K)& log(g) &  v$_{\rm t}$ (km/s) & U & B & V & I$_{\rm C}$ & J & H & K \\
\hline
\endhead
\hline
\endfoot
19925 & 245.81007350 & -26.60153694 & 4050 & 1.20 & 1.67 & 14.83 & 12.76 & 11.04 &  8.97 &  7.59 &  6.63 &  6.38\\
20766 & 245.80811060 & -26.55680649 & 4400 & 1.80 & 1.45 & 14.89 & 13.53 & 12.10 & 10.33 &  9.07 &  8.24 &  8.07\\
21191 & 245.81900040 & -26.53602484 & 4270 & 1.60 & 1.60 & 14.79 & 13.21 & 11.70 &  9.85 &  8.50 &  7.67 &  7.45\\
21728 & 245.81082240 & -26.51303931 & 4525 & 2.00 & 1.42 & 15.05 & 13.86 & 12.51 & 10.78 &  9.50 &  8.75 &  8.51\\
22089 & 245.82041320 & -26.49663964 & 4700 & 2.28 & 1.36 & 15.28 & 14.36 & 13.14 & 11.54 & 10.34 &  9.61 &  9.47\\
24590 & 245.91667710 & -26.65088182 & 4850 & 2.66 & 1.35 & 16.89 & 14.67 & 13.44 & 11.88 & 10.79 & 10.11 &  9.94\\
25709 & 245.92048770 & -26.62319472 & 4680 & 2.20 & 1.38 & 15.11 & 14.24 & 12.96 & 11.31 & 10.16 &  9.45 &  9.28\\
26471 & 245.88130650 & -26.60513670 & 4800 & 2.40 & 1.28 & 15.38 & 14.70 & 13.49 & 11.87 & 10.73 & 10.05 &  9.90\\
26794 & 245.90940720 & -26.59866016 & 4800 & 2.45 & 1.44 & 15.69 & 14.94 & 13.71 & 12.08 & 10.92 & 10.25 & 10.07\\
27448 & 245.95330690 & -26.58586889 & 4310 & 1.57 & 1.58 &   -   & 13.24 & 11.73 &  9.86 &  8.57 &  7.71 &  7.53\\
28103 & 245.84921830 & -26.57493123 & 3860 & 0.50 & 1.62 & 14.75 & 12.52 & 10.71 &  8.51 &  7.01 &  6.03 &  5.75\\
28356 & 245.97324550 & -26.57086633 & 4600 & 2.22 & 1.53 & 14.93 & 13.92 & 12.57 & 10.87 &  9.67 &  8.92 &  8.72\\
28707 & 245.93837450 & -26.56579907 & 4880 & 2.74 & 1.31 & 15.59 & 14.84 & 13.61 & 11.99 & 10.80 & 10.14 &  9.95\\
28797 & 245.97434580 & -26.56458317 & 4640 & 2.35 & 1.36 & 15.13 & 14.22 & 12.93 & 11.30 & 10.12 &  9.43 &  9.21\\
28847 & 245.90471930 & -26.56390736 & 4780 & 2.40 & 1.27 & 15.43 & 14.76 & 13.54 & 11.88 & 10.67 &  9.98 &  9.77\\
28977 & 245.94551720 & -26.56218478 & 4680 & 2.33 & 1.40 & 15.22 & 14.31 & 13.02 & 11.36 & 10.14 &  9.44 &  9.26\\
29027 & 245.84810750 & -26.56146317 & 4720 & 2.40 & 1.41 & 15.18 & 14.40 & 13.16 & 11.48 & 10.29 &  9.58 &  9.39\\
29065 & 245.94135950 & -26.56081120 & 4650 & 2.10 & 1.41 & 15.08 & 14.27 & 12.98 & 11.31 & 10.10 &  9.42 &  9.24\\
29171 & 245.89969150 & -26.55958664 & 4880 & 2.64 & 1.26 & 15.72 & 14.95 & 13.73 & 12.09 & 10.88 & 10.16 & 10.00\\
29222 & 245.84606690 & -26.55896241 & 4720 & 2.50 & 1.36 & 15.28 & 14.42 & 13.17 & 11.52 & 10.32 &  9.59 &  9.41\\
29272 & 245.86286750 & -26.55825152 & 4780 & 2.50 & 1.26 & 15.32 & 14.64 & 13.42 & 11.76 & 10.54 &  9.86 &  9.67\\
29282 & 245.85544600 & -26.55811142 & 4650 & 2.30 & 1.42 & 15.17 & 14.17 & 12.88 & 11.18 &  9.91 &  9.20 &  8.98\\
29397 & 245.87588880 & -26.55671383 & 4600 & 1.50 & 1.78 & 14.60 & 13.53 & 12.20 & 10.47 &  9.25 &  8.54 &  8.32\\
29545 & 245.92749790 & -26.55485385 & 4880 & 2.61 & 1.18 & 15.53 & 14.96 & 13.79 & 12.23 & 11.08 & 10.43 & 10.24\\
29598 & 245.88908940 & -26.55414514 & 4840 & 2.50 & 1.40 & 15.65 & 14.89 & 13.68 & 12.05 & 10.86 & 10.19 & 10.01\\
29693 & 245.88094110 & -26.55307864 & 4360 & 1.10 & 1.88 & 14.67 & 13.27 & 11.80 &  9.96 &  8.63 &  7.79 &  7.64\\
29848 & 245.89219050 & -26.55108648 & 4780 & 2.52 & 1.24 & 15.48 & 14.75 & 13.52 & 11.87 & 10.69 & 10.00 &  9.83\\
30209 & 245.89243520 & -26.54681482 & 4880 & 2.62 & 1.32 & 15.52 & 14.87 & 13.66 & 12.05 & 10.84 & 10.15 &  9.99\\
30345 & 245.94994400 & -26.54524113 & 4850 & 2.73 & 1.31 & 15.64 & 15.00 & 13.84 & 12.31 & 11.19 & 10.55 & 10.39\\
30450 & 245.86144840 & -26.54411333 & 4760 & 2.53 & 1.35 & 15.44 & 14.59 & 13.34 & 11.69 & 10.47 &  9.78 &  9.59\\
30452 & 245.88768830 & -26.54411091 & 4830 & 2.56 & 1.25 & 15.72 & 15.07 & 13.88 & 12.26 & 11.07 & 10.38 & 10.22\\
30549 & 245.88815200 & -26.54294543 & 4830 & 2.52 & 1.28 & 15.52 & 14.86 & 13.66 & 12.05 & 10.76 & 10.13 &  9.99\\
30598 & 245.89661740 & -26.54235225 & 4360 & 1.75 & 1.47 & 14.72 & 13.49 & 12.05 & 10.23 &  8.83 &  8.03 &  7.75\\
30653 & 245.88793300 & -26.54184062 & 4660 & 2.30 & 1.25 & 15.23 & 14.40 & 13.14 & 11.50 & 10.29 &  9.56 &  9.38\\
30675 & 245.90236690 & -26.54157800 & 4830 & 2.58 & 1.35 & 15.52 & 14.77 & 13.58 & 12.00 &   -   &   -   &   -	\\
30711 & 245.85156170 & -26.54123206 & 4560 & 2.25 & 1.46 & 15.04 & 13.86 & 12.49 & 10.73 &  9.44 &  8.66 &  8.46\\
30719 & 245.90310860 & -26.54116153 & 4810 & 2.65 & 1.24 & 15.79 & 15.13 & 13.97 & 12.43 &   -   &   -   &   -	\\
30751 & 245.94302430 & -26.54089574 & 4430 & 1.78 & 1.47 & 14.83 & 13.61 & 12.20 & 10.45 &  9.19 &  8.36 &  8.16\\
30924 & 245.89624320 & -26.53883396 & 4810 & 2.60 & 1.28 & 15.48 & 14.79 & 13.61 & 12.03 &  9.55 &  8.81 &  8.40\\
30933 & 245.95895830 & -26.53873652 & 4800 & 2.63 & 1.30 & 15.46 & 14.70 & 13.49 & 11.92 & 10.77 & 10.12 &  9.91\\
31015 & 245.89710880 & -26.53795346 & 4800 & 2.47 & 1.37 & 14.91 & 14.21 & 13.00 & 11.38 & 10.17 &  9.48 &  9.29\\
31306 & 245.92607560 & -26.53505913 & 4900 & 2.87 & 1.33 & 15.79 & 15.12 & 13.97 & 12.44 & 11.32 & 10.64 & 10.50\\
31376 & 245.91219110 & -26.53432365 & 4800 & 2.59 & 1.36 & 15.37 & 14.52 & 13.29 & 11.68 & 10.46 &  9.79 &  9.59\\
31532 & 245.89561610 & -26.53290071 & 4770 & 2.60 & 1.21 & 15.28 & 14.53 & 13.33 & 11.76 & 10.54 &  9.85 &  9.72\\
31665 & 245.93007130 & -26.53164468 & 4650 & 2.17 & 1.34 & 14.94 & 14.12 & 12.85 & 11.18 &  9.96 &  9.25 &  9.04\\
31803 & 245.90925770 & -26.53023616 & 4850 & 2.60 & 1.34 & 15.77 & 15.14 & 13.99 & 12.46 & 11.31 & 10.64 & 10.51\\
31845 & 245.85370350 & -26.52982260 & 4700 & 2.42 & 1.31 & 15.23 & 14.34 & 13.09 & 11.44 & 10.16 &  9.49 &  9.29\\
32055 & 245.91709840 & -26.52773869 & 4300 & 1.57 & 1.52 & 14.71 & 13.36 & 11.88 & 10.06 &  8.71 &  7.90 &  7.66\\
32121 & 245.83812920 & -26.52713080 & 4840 & 2.58 & 1.33 & 15.49 & 14.67 & 13.45 & 11.84 & 10.64 &  9.93 &  9.75\\
32151 & 245.83708630 & -26.52688433 & 4770 & 2.43 & 1.38 & 15.41 & 14.59 & 13.38 & 11.75 & 10.54 &  9.84 &  9.70\\
32317 & 245.86225000 & -26.52536111 & 4510 & 1.88 & 1.43 & 15.06 & 13.93 & 12.56 &   -   &  9.51 &  8.76 &  8.57\\
32347 & 245.92079170 & -26.52505556 & 4640 & 2.22 & 1.38 &   -   &   -   &   -   &   -   &  9.92 &  9.20 &  9.00\\
32583 & 245.89985980 & -26.52267062 & 4850 & 2.54 & 1.34 & 15.46 & 14.80 & 13.61 & 12.03 & 10.80 & 10.16 &  9.95\\
32627 & 245.97840770 & -26.52219757 & 4750 & 2.42 & 1.30 & 15.01 & 14.34 & 13.14 & 11.59 & 10.48 &  9.79 &  9.61\\
32700 & 245.91714040 & -26.52145861 & 4560 & 2.12 & 1.36 & 14.88 & 13.86 & 12.52 & 10.80 &  9.54 &  8.81 &  8.57\\
32724 & 245.92395710 & -26.52127838 & 4850 & 2.73 & 1.30 & 15.53 & 14.82 & 13.62 & 12.04 & 10.88 & 10.20 & 10.04\\
32782 & 245.89106390 & -26.52065353 & 4880 & 2.60 & 1.24 & 15.81 & 15.10 & 13.94 & 12.38 &   -   &   -   &   -	\\
32871 & 245.85272450 & -26.51984366 & 4770 & 2.48 & 1.24 & 15.65 & 14.92 & 13.72 & 12.09 & 10.88 & 10.22 & 10.01\\
32874 & 245.91452240 & -26.51985129 & 4600 & 2.04 & 1.38 & 14.82 & 13.99 & 12.72 & 11.07 &  9.86 &  9.13 &  8.93\\
32933 & 245.83915100 & -26.51926550 & 4430 & 1.42 & 1.78 & 14.54 & 13.29 & 11.87 & 10.08 &  8.81 &  8.02 &  7.79\\
32968 & 245.89176500 & -26.51884367 & 4630 & 2.17 & 1.30 & 15.06 & 14.21 & 12.96 & 11.32 & 10.07 &  9.41 &  9.15\\
32988 & 245.90955560 & -26.51865553 & 4850 & 2.63 & 1.22 & 15.43 & 14.88 & 13.75 & 12.22 & 11.08 & 10.40 & 10.22\\
33069 & 245.94851290 & -26.51783666 & 4940 & 3.05 & 1.36 & 15.45 & 14.71 & 13.53 & 11.99 & 10.86 & 10.19 & 10.02\\
33195 & 245.86426280 & -26.51654913 & 4620 & 2.38 & 1.26 & 15.30 & 14.39 & 13.12 & 11.45 & 10.23 &  9.50 &  9.36\\
33414 & 245.84207930 & -26.51444711 & 4840 & 2.51 & 1.28 & 15.51 & 14.84 & 13.67 & 12.08 & 10.90 & 10.21 & 10.09\\
33617 & 245.86252960 & -26.51233938 & 4720 & 2.35 & 1.26 & 15.38 & 14.57 & 13.36 & 11.76 & 10.59 &  9.86 &  9.68\\
33629 & 245.83816250 & -26.51221901 & 4930 & 2.80 & 1.33 & 15.77 & 15.06 & 13.90 & 12.34 & 11.17 & 10.48 & 10.36\\
33683 & 245.89914420 & -26.51176320 & 4800 & 2.57 & 1.18 & 15.29 & 14.66 & 13.48 & 11.91 & 10.78 & 10.09 &  9.89\\
33788 & 245.91070220 & -26.51060047 & 4700 & 2.37 & 1.33 & 15.15 & 14.24 & 12.99 & 11.38 & 10.16 &  9.44 &  9.27\\
33900 & 245.84591550 & -26.50933155 & 4770 & 2.48 & 1.27 & 15.48 & 14.69 & 13.51 & 11.93 & 10.67 & 10.03 &  9.92\\
33946 & 245.84147130 & -26.50887807 & 4800 & 2.62 & 1.15 & 15.63 & 14.89 & 13.71 & 12.14 & 10.95 & 10.29 & 10.15\\
34006 & 245.87135820 & -26.50826984 & 4320 & 1.67 & 1.61 & 14.92 & 13.37 & 11.87 &  9.97 &  8.59 &  7.81 &  7.51\\
34130 & 245.88104790 & -26.50687941 & 4550 & 2.08 & 1.40 & 14.98 & 13.87 & 12.54 & 10.82 &  9.58 &  8.80 &  8.60\\
34167 & 245.90419020 & -26.50638230 & 4950 & 2.60 & 1.40 & 15.63 & 15.08 & 13.95 & 12.44 & 11.31 & 10.67 & 10.52\\
34240 & 245.90241860 & -26.50557893 & 4470 & 1.95 & 1.41 & 14.86 & 13.74 & 12.37 & 10.63 &  9.35 &  8.62 &  8.43\\
34502 & 245.92046770 & -26.50260567 & 4860 & 2.70 & 1.33 & 15.49 & 14.78 & 13.62 & 12.09 & 10.96 & 10.26 & 10.15\\
34579 & 245.93359340 & -26.50170174 & 4330 & 1.59 & 1.49 & 14.67 & 13.27 & 11.83 & 10.07 &  8.77 &  7.91 &  7.71\\
34726 & 245.85009660 & -26.50013473 & 4600 & 2.24 & 1.35 & 15.17 & 14.14 & 12.86 & 11.16 &  9.89 &  9.13 &  8.99\\
35022 & 245.94949140 & -26.49673818 & 4850 & 2.51 & 1.36 & 15.50 & 14.79 & 13.62 & 12.10 & 10.96 & 10.31 & 10.11\\
35061 & 245.92398380 & -26.49632071 & 4860 & 2.67 & 1.23 & 15.68 & 15.02 & 13.85 & 12.31 & 11.15 & 10.48 & 10.34\\
35455 & 245.92832620 & -26.49099726 & 4600 & 2.10 & 1.29 & 14.89 & 13.99 & 12.71 & 11.05 &  9.83 &  9.08 &  8.92\\
35487 & 245.95940170 & -26.49051723 & 4850 & 2.67 & 1.24 & 15.33 & 14.61 & 13.45 & 11.95 & 10.82 & 10.18 &  9.99\\
35508 & 245.85414850 & -26.49026951 & 4780 & 2.48 & 1.18 & 15.66 & 14.96 & 13.80 & 12.23 & 11.05 & 10.37 & 10.18\\
35571 & 245.88484770 & -26.48951670 & 4880 & 2.79 & 1.10 & 15.66 & 15.07 & 13.93 & 12.39 & 11.25 & 10.61 & 10.42\\
35627 & 245.95411440 & -26.48867681 & 4830 & 2.37 & 1.20 & 15.48 & 14.94 & 13.80 & 12.30 & 11.14 & 10.50 & 10.34\\
35688 & 245.91086870 & -26.48777246 & 4720 & 2.25 & 1.33 & 15.32 & 14.48 & 13.27 & 11.68 & 10.46 &  9.75 &  9.57\\
35774 & 245.87296550 & -26.48657771 & 4450 & 1.92 & 1.44 & 14.88 & 13.68 & 12.32 & 10.59 &  9.32 &  8.57 &  8.35\\
36215 & 245.91658540 & -26.48028356 & 4300 & 1.59 & 1.53 & 14.74 & 13.29 & 11.80 &  9.96 &  8.61 &  7.82 &  7.57\\
36356 & 245.90812220 & -26.47815339 & 4820 & 2.66 & 1.26 & 15.52 & 14.76 & 13.57 & 12.01 & 10.82 & 10.17 &  9.97\\
36929 & 245.92953710 & -26.46873630 & 4820 & 2.55 & 1.28 & 15.50 & 14.75 & 13.59 & 12.07 & 10.93 & 10.25 & 10.07\\
36942 & 245.97083610 & -26.46848900 & 4800 & 2.66 & 1.23 & 15.43 & 14.74 & 13.56 & 12.04 & 10.90 & 10.23 & 10.08\\
37215 & 245.89085360 & -26.46381984 & 4790 & 2.50 & 1.21 & 15.61 & 14.90 & 13.72 & 12.15 & 10.94 & 10.26 & 10.11\\
38075 & 245.84038430 & -26.44634617 & 4800 & 2.54 & 1.25 & 15.61 & 14.83 & 13.65 & 12.05 & 10.82 & 10.18 &  9.99\\
38383 & 245.84437580 & -26.43947709 & 4590 & 1.87 & 1.45 & 15.02 & 14.03 & 12.74 & 11.02 &  9.74 &  8.99 &  8.79\\
38399 & 245.88487200 & -26.43904923 & 4730 & 2.35 & 1.38 & 15.23 & 14.37 & 13.14 & 11.51 & 10.29 &  9.56 &  9.39\\
38896 & 245.91189860 & -26.42854610 & 4760 & 2.53 & 1.31 & 15.26 & 14.36 & 13.14 & 11.55 & 10.33 &  9.64 &  9.46\\
42490 & 246.00378460 & -26.59808973 & 4570 & 2.08 & 1.41 & 14.95 & 13.94 & 12.56 & 10.84 &  9.71 &  8.95 &  8.76\\
42620 & 245.99313500 & -26.59417329 & 4600 & 2.05 & 1.37 & 14.84 & 13.91 & 12.59 & 10.94 &  9.85 &  9.11 &  8.97\\
43370 & 246.03196790 & -26.57365632 & 4920 & 2.80 & 1.32 & 15.63 & 15.03 & 13.85 & 12.31 & 11.29 & 10.66 & 10.50\\
44243 & 246.02382800 & -26.55029031 & 4860 & 2.80 & 1.17 & 15.51 & 14.91 & 13.75 & 12.24 & 11.23 & 10.59 & 10.42\\
44595 & 246.06621120 & -26.54096589 & 4750 & 2.40 & 1.40 & 15.02 & 14.23 & 13.00 & 11.45 & 10.42 &  9.75 &  9.58\\
44616 & 245.98944100 & -26.54022147 & 4620 & 2.20 & 1.44 & 14.99 & 13.99 & 12.65 & 10.99 &  9.82 &  9.11 &  8.89\\
45163 & 246.00502890 & -26.52572879 & 4770 & 2.40 & 1.26 & 15.27 & 14.59 & 13.40 & 11.88 & 10.81 & 10.17 &  9.98\\
45895 & 246.01787760 & -26.50775258 & 4720 & 2.25 & 1.34 & 14.93 & 14.22 & 12.99 & 11.43 & 10.38 &  9.71 &  9.52\\
 5359 & 245.91226500 & -26.36929347 & 4800 & 2.44 & 1.28 & 15.60 & 14.78 & 13.58 & 12.01 & 10.87 & 10.21 & 10.05\\
\end{longtable}
}

%\end{document}

\clearpage

\longtab{9}{
\begin{longtable}{lcccccccccccc}
\caption{\label{Tab9} Chemical abundances of the analised stars.}\\
\hline
\hline
\scriptsize{ID} & \scriptsize{[FeI/H]} & \scriptsize{[OI/Fe]} & \scriptsize{[NaI/Fe]$_{NLTE}$} & \scriptsize{[MgI/Fe]$_{NLTE}$} & \scriptsize{[AlI/Fe]} & \scriptsize{[SiI/Fe]} & \scriptsize{[CaI/Fe]} & \scriptsize{[TiI/Fe]} & \scriptsize{[TiII/Fe]} & \scriptsize{[CrI/Fe]} & \scriptsize{[NiI/Fe]} & \scriptsize{[BaII/Fe]}\\
\hline
\endfirsthead
\caption{continued.}\\
\hline\hline
\scriptsize{ID} & \scriptsize{[FeI/H]} & \scriptsize{[OI/Fe]} & \scriptsize{[NaI/Fe]$_{NLTE}$} & \scriptsize{[MgI/Fe]$_{NLTE}$} & \scriptsize{[AlI/Fe]} & \scriptsize{[SiI/Fe]} & \scriptsize{[CaI/Fe]} & \scriptsize{[TiI/Fe]} & \scriptsize{[TiII/Fe]} & \scriptsize{[CrI/Fe]} & \scriptsize{[NiI/Fe]} & \scriptsize{[BaII/Fe]}\\
\hline
\endhead
\hline
\endfoot
19925 & $-$1.02 & 0.28 & 0.51   &  0.43 & 0.68 & 0.50 & 0.19 &  0.48 & 0.29 & $-$0.04 &    0.09 &  0.64 \\
20766 & $-$1.05 & 0.31 & 0.53   &  0.49 & 0.62 & 0.49 & 0.26 &  0.31 & 0.39 & $-$0.01 &    0.04 &  0.45 \\ 
21191 & $-$1.06 & 0.34 & 0.51   &  0.55 & 0.59 & 0.54 & 0.24 &  0.31 & 0.32 & $-$0.00 &    0.03 &  0.53 \\ 
21728 & $-$1.06 & 0.31 & 0.37   &  0.52 & 0.68 & 0.47 & 0.28 &  0.32 & 0.30 & $-$0.04 &    0.03 &  0.48 \\ 
22089 & $-$1.06 & 0.35 & 0.50   &  0.30 & --   & 0.53 & 0.31 &  0.25 & 0.32 & $-$0.15 &    0.03 &  0.46 \\ 
24590 & $-$1.07 & 0.50 & 0.30   &  0.61 & --   & 0.50 & 0.27 &  0.29 & 0.31 & $-$0.09 &    0.04 &  0.33 \\ 
25709 & $-$1.13 & 0.44 & 0.34   &  0.54 & 0.59 & 0.53 & 0.35 &  0.26 & 0.33 & $-$0.02 & $-$0.01 &  0.35 \\ 
26471 & $-$1.10 & 0.41 & 0.09   &  0.59 & 0.58 & 0.45 & 0.30 &  0.23 & 0.31 &    0.00 & $-$0.01 &  0.28 \\ 
26794 & $-$1.17 & --   & 0.36   &  0.53 & --   & 0.50 & 0.34 &  0.34 & 0.39 & $-$0.01 &    0.03 &  0.33 \\ 
27448 & $-$1.12 & 0.51 & 0.11   &  0.50 & 0.42 & 0.45 & 0.29 &  0.39 & 0.36 & $-$0.02 &    0.03 &  0.35 \\ 
28103 & $-$1.08 & 0.50 & 0.17   &  0.41 & 0.33 & 0.55 & 0.16 &  0.48 & 0.26 & $-$0.11 &    0.01 &  0.26 \\ 
28356 & $-$1.14 & 0.44 & 0.37   &  0.55 & 0.57 & 0.53 & 0.28 &  0.29 & 0.37 &    0.04 &    0.05 &  0.39 \\ 
28707 & $-$1.03 & --   & 0.22   &  0.40 & --   & 0.43 & 0.20 &  0.27 & 0.26 & $-$0.06 &    0.02 &  0.45 \\ 
28797 & $-$1.12 & 0.35 & 0.44   &  0.55 & 0.57 & 0.57 & 0.26 &  0.28 & 0.44 & $-$0.12 &    0.01 &  0.53 \\ 
28847 & $-$1.16 & 0.52 & 0.08   &  0.52 & --   & 0.51 & 0.32 &  0.31 & 0.30 & $-$0.09 & $-$0.04 &  0.46 \\ 
28977 & $-$1.14 & 0.39 & 0.40   &  0.51 & 0.65 & 0.57 & 0.30 &  0.31 & 0.35 & $-$0.02 &    0.05 &  0.42 \\ 
29027 & $-$1.10 & 0.51 & 0.02   &  0.48 & --   & 0.44 & 0.27 &  0.27 & 0.35 & $-$0.01 & $-$0.03 &  0.34 \\ 
29065 & $-$1.12 & 0.45 & 0.17   &  0.61 & --   & 0.48 & 0.27 &  0.21 & 0.35 & $-$0.06 & $-$0.01 &  0.35 \\ 
29171 & $-$0.99 & 0.22 & 0.41   &  0.43 & --   & 0.49 & 0.32 &  0.32 & 0.31 & $-$0.04 & $-$0.04 &  0.36 \\ 
29222 & $-$1.04 & 0.36 & 0.24   &  0.44 & 0.44 & 0.47 & 0.25 &  0.27 & 0.58 & $-$0.06 &    0.05 &  0.39 \\ 
29272 & $-$1.11 & 0.53 & 0.05   &  0.47 & --   & 0.43 & 0.25 &  0.30 & 0.33 &    0.01 &    0.02 &  0.35 \\ 
29282 & $-$1.06 & 0.28 & 0.42   &  0.50 & 0.61 & 0.49 & 0.25 &  0.23 & 0.34 & $-$0.07 &    0.02 &  0.27 \\ 
29397 & $-$1.12 & 0.24 & 0.46   &  0.55 & 0.51 & 0.53 & 0.20 &  0.20 & 0.15 & $-$0.11 & $-$0.03 &  0.43 \\ 
29545 & $-$1.06 & 0.47 &$-$0.02 &  0.44 & 0.15 & 0.36 & 0.29 &  0.29 & 0.33 & $-$0.07 &    0.00 &  0.20 \\ 
29598 & $-$1.06 & 0.39 & 0.40   &  0.48 & --   & 0.50 & 0.34 &  0.26 & 0.42 & $-$0.04 &    0.02 &  0.28 \\ 
29693 & $-$1.19 & 0.26 & 0.50   &  0.60 & 0.64 & 0.55 & 0.14 &  0.23 & 0.28 & $-$0.17 &    0.00 &  0.36 \\ 
29848 & $-$1.05 & 0.54 & 0.09   &  0.54 & 0.64 & 0.48 & 0.29 &  0.32 & 0.34 & $-$0.01 & $-$0.01 &  0.55 \\ 
30209 & $-$0.99 & 0.42 &$-$0.05 &  0.46 & --   & 0.42 & 0.30 &  0.31 & 0.46 &    0.03 &    0.05 &  0.31 \\ 
30345 & $-$1.06 & 0.32 & 0.43   &  0.44 & --   & 0.53 & 0.28 &  0.28 & 0.52 & $-$0.05 &    0.07 &  0.36 \\ 
30450 & $-$1.00 & 0.40 & 0.40   &  0.52 & 0.51 & 0.49 & 0.26 &  0.29 & 0.32 &    0.11 &    0.05 &  0.45 \\ 
30452 & $-$1.00 & 0.40 & 0.06   &  0.50 & 0.53 & 0.46 & 0.27 &  0.29 & 0.36 &    0.02 &    0.00 &  0.53 \\ 
30549 & $-$1.09 & 0.45 & 0.13   &  0.52 & --   & 0.48 & 0.34 &  0.31 & 0.35 &    0.00 &    0.06 &  0.40 \\ 
30598 & $-$1.07 & 0.43 & 0.05   &  0.54 & 0.37 & 0.54 & 0.25 &  0.33 & 0.41 &    0.01 &    0.05 &  0.52 \\ 
30653 & $-$1.06 & 0.47 & 0.15   &  0.47 & 0.46 & 0.51 & 0.27 &  0.21 & 0.32 & $-$0.06 & $-$0.01 &  0.45 \\ 
30675 & $-$1.07 & 0.36 & 0.41   &  0.55 & 0.65 & 0.50 & 0.36 &  0.32 & 0.42 & $-$0.02 &    0.07 &  0.35 \\ 
30711 & $-$1.01 & 0.37 & 0.32   &  0.47 & 0.53 & 0.45 & 0.22 &  0.31 & 0.41 & $-$0.07 &    0.02 &  0.46 \\ 
30719 & $-$1.19 & 0.44 & 0.42   &  0.50 & 0.68 & 0.46 & 0.33 &  0.27 & 0.48 & $-$0.09 &    0.02 &  0.45 \\ 
30751 & $-$1.09 & 0.40 & 0.32   &  0.58 & 0.54 & 0.48 & 0.28 &  0.38 & 0.33 &    0.10 &    0.03 &  0.42 \\ 
30924 & $-$1.09 & 0.48 & 0.20   &  0.47 & 0.48 & 0.50 & 0.28 &  0.30 & 0.41 & $-$0.05 &    0.02 &  0.40 \\ 
30933 & $-$1.07 & 0.44 & 0.44   &  0.44 & --   & 0.58 & 0.30 &  0.28 & 0.34 & $-$0.08 &    0.02 &  0.46 \\ 
31015 & $-$1.07 & 0.50 & 0.00   &  0.51 & 0.36 & 0.42 & 0.33 &  0.32 & 0.40 &    0.06 & $-$0.01 &  0.32 \\ 
31306 & $-$1.11 & --   & 0.40   &  0.41 & 0.53 & 0.45 & 0.29 &  0.33 & 0.38 &    0.03 &    0.04 &  0.36 \\ 
31376 & $-$1.00 & 0.33 & 0.43   &  0.46 & 0.59 & 0.43 & 0.29 &  0.29 & 0.34 & $-$0.01 &    0.03 &  0.40 \\ 
31532 & $-$1.03 & 0.47 & 0.38   &  0.44 & 0.55 & 0.47 & 0.26 &  0.30 & 0.39 & $-$0.08 &    0.00 &  0.49 \\ 
31665 & $-$1.07 & 0.48 & 0.02   &  0.49 & 0.29 & 0.46 & 0.31 &  0.26 & 0.35 & $-$0.06 &    0.01 &  0.29 \\ 
31803 & $-$1.13 & 0.38 & 0.25   &  0.44 & 0.49 & 0.51 & 0.32 &  0.31 & 0.28 & $-$0.10 & $-$0.01 &  0.29 \\ 
31845 & $-$1.06 & 0.27 & 0.38   &  0.41 & 0.67 & 0.48 & 0.27 &  0.27 & 0.35 & $-$0.09 &    0.02 &  0.47 \\ 
32055 & $-$1.12 & 0.31 & 0.40   &  0.55 & 0.59 & 0.52 & 0.27 &  0.39 & 0.28 &    0.03 &    0.00 &  0.45 \\ 
32121 & $-$0.93 & --   & 0.35   &  0.44 & 0.47 & 0.56 & 0.29 &  0.32 & 0.37 & $-$0.08 &    0.02 &  0.38 \\ 
32151 & $-$1.07 & --   & 0.39   &  0.42 & --   & 0.49 & 0.29 &  0.26 & 0.34 & $-$0.04 & $-$0.01 &  0.33 \\ 
32317 & $-$1.07 & 0.37 & 0.37   &  0.54 & --   & 0.43 & 0.31 &  0.33 & 0.34 &    0.08 &    0.03 &  0.36 \\ 
32347 & $-$1.09 & 0.40 & 0.23   &  0.51 & 0.57 & 0.49 & 0.27 &  0.29 & 0.32 & $-$0.06 &    0.03 &  0.43 \\ 
32583 & $-$1.08 & 0.46 & 0.07   &  0.46 & 0.50 & 0.39 & 0.31 &  0.32 & 0.34 & $-$0.07 &    0.00 &  0.29 \\ 
32627 & $-$1.10 & 0.50 &$-$0.03 &  0.51 & 0.36 & 0.34 & 0.29 &  0.26 & 0.38 & $-$0.01 & $-$0.00 &  0.36 \\ 
32700 & $-$1.02 & 0.49 & 0.06   &  0.61 & 0.56 & 0.47 & 0.22 &  0.37 & 0.39 &    0.06 &    0.00 &  0.52 \\
32724 & $-$1.03 & 0.51 & 0.11   &  0.48 & 0.53 & 0.39 & 0.30 &  0.34 & 0.39 & $-$0.04 &    0.00 &  0.49 \\ 
32782 & $-$1.05 & 0.29 & 0.46   &  0.45 & 0.58 & 0.41 & 0.34 &  0.26 & 0.27 & $-$0.07 & $-$0.03 &  0.48 \\ 
32871 & $-$1.01 & 0.40 & 0.06   &  0.62 & 0.81 & 0.54 & 0.29 &  0.31 & 0.34 & $-$0.05 &    0.01 &  0.65 \\ 
32874 & $-$1.14 & 0.41 & 0.11   &  0.55 & 0.52 & 0.42 & 0.26 &  0.27 & 0.32 &    0.02 & $-$0.02 &  0.39 \\ 
32933 & $-$1.13 & 0.48 & 0.02   &  0.56 & --   & 0.54 & 0.20 &  0.29 & 0.22 & $-$0.14 &    0.04 &  0.53 \\ 
32968 & $-$1.13 & 0.41 & 0.24   &  0.54 & 0.50 & 0.49 & 0.30 &  0.25 & 0.42 & $-$0.03 &    0.01 &  0.48 \\ 
32988 & $-$1.09 & 0.51 & 0.02   &  0.44 & --   & 0.50 & 0.29 &  0.29 & 0.33 & $-$0.03 & $-$0.01 &  --   \\ 
33069 & $-$0.92 & 0.39 & 0.23   &  0.37 & 0.59 & 0.43 & 0.20 &  0.34 & 0.42 & $-$0.06 &    0.03 &  0.42 \\ 
33195 & $-$1.03 & 0.48 & 0.09   &  0.60 & 0.66 & 0.56 & 0.27 &  0.32 & 0.34 & $-$0.02 &    0.04 &  --   \\
33414 & $-$1.05 & 0.52 & 0.21   &  0.49 & 0.30 & 0.48 & 0.32 &  0.25 & 0.26 & $-$0.05 &    0.00 &  0.36 \\ 
33617 & $-$1.09 & 0.29 & 0.37   &  0.54 & 0.58 & 0.51 & 0.32 &  0.26 & 0.31 & $-$0.05 &    0.02 &  0.51 \\ 
33629 & $-$0.98 & --   & 0.31   &  0.39 & 0.68 & 0.55 & 0.28 &  0.33 & 0.37 & $-$0.02 &    0.05 &  0.34 \\ 
33683 & $-$1.05 & 0.42 & 0.00   &  0.51 & 0.51 & 0.42 & 0.27 &  0.27 & 0.39 &    0.03 & $-$0.03 &  0.56 \\ 
33788 & $-$1.02 & 0.30 & 0.37   &  0.47 & 0.50 & 0.43 & 0.28 &  0.28 & 0.33 & $-$0.02 &    0.00 &  0.48 \\ 
33900 & $-$1.06 & 0.30 & 0.47   &  0.49 & 0.55 & 0.46 & 0.32 &  0.30 & 0.39 & $-$0.05 &    0.01 &  0.42 \\ 
33946 & $-$1.03 & 0.23 & 0.32   &  0.42 & 0.60 & 0.45 & 0.30 &  0.29 & 0.33 & $-$0.03 &    0.03 &  0.50 \\ 
34006 & $-$1.06 & 0.25 & 0.44   &  0.52 & 0.67 & 0.48 & 0.26 &  0.37 & 0.27 & $-$0.08 &    0.04 &  0.42 \\ 
34130 & $-$1.09 & 0.33 & 0.43   &  0.53 & 0.59 & 0.51 & 0.29 &  0.26 & 0.34 & $-$0.04 & $-$0.01 &  0.47 \\ 
34167 & $-$1.10 & --   & 0.13   &  0.56 & 0.48 & 0.35 & 0.31 &  0.25 & 0.23 &    0.01 & $-$0.02 &  0.04 \\ 
34240 & $-$1.10 & 0.47 & 0.08   &  0.55 & 0.55 & 0.49 & 0.25 &  0.35 & 0.30 &    0.04 &    0.03 &  0.47 \\ 
34502 & $-$1.08 & 0.35 & 0.34   &  0.47 & 0.49 & 0.47 & 0.28 &  0.33 & 0.38 & $-$0.06 &    0.02 &  0.34 \\ 
34579 & $-$1.08 & 0.25 & 0.46   &  0.52 & 0.61 & 0.50 & 0.28 &  0.38 & 0.32 & $-$0.06 &    0.02 &  0.46 \\ 
34726 & $-$1.01 & 0.37 & 0.42   &  0.49 & 0.60 & 0.46 & 0.28 &  0.31 & 0.34 & $-$0.06 &    0.03 &  0.40 \\ 
35022 & $-$1.08 & 0.30 & 0.41   &  0.52 & 0.58 & 0.50 & 0.33 &  0.32 & 0.22 & $-$0.05 &    0.02 &  0.34 \\ 
35061 & $-$0.99 & 0.50 & 0.06   &  0.50 & 0.52 & 0.40 & 0.24 &  0.24 & 0.33 & $-$0.06 & $-$0.01 &  0.39 \\ 
35455 & $-$1.06 & 0.45 & 0.01   &  0.49 & 0.49 & 0.44 & 0.24 &  0.23 & 0.37 & $-$0.01 & $-$0.02 &  0.53 \\ 
35487 & $-$1.00 & 0.36 & 0.30   &  0.42 & 0.55 & 0.44 & 0.27 &  0.32 & 0.43 & $-$0.05 & $-$0.01 &  0.51 \\ 
35508 & $-$1.05 & 0.25 & 0.38   &  0.41 & 0.45 & 0.45 & 0.27 &  0.22 & 0.27 & $-$0.11 & $-$0.05 &  0.46 \\ 
35571 & $-$0.99 & 0.45 &$-$0.02 &  0.46 & 0.52 & 0.45 & 0.27 &  0.24 & 0.34 & $-$0.09 & $-$0.04 &  0.55 \\ 
35627 & $-$1.08 & --   & 0.04   &  0.56 & 0.51 & 0.38 & 0.33 &  0.26 & 0.38 & $-$0.09 &    0.00 &  0.29 \\ 
35688 & $-$1.11 & 0.21 & 0.40   &  0.53 & 0.59 & 0.47 & 0.33 &  0.23 & 0.34 & $-$0.04 &    0.02 &  0.40 \\ 
35774 & $-$1.10 & 0.43 & 0.24   &  0.54 & 0.51 & 0.53 & 0.26 &  0.36 & 0.31 & $-$0.03 &    0.03 &  0.43 \\ 
36215 & $-$1.11 & 0.48 & 0.18   &  0.52 & 0.53 & 0.53 & 0.26 &  0.38 & 0.34 & $-$0.04 &    0.02 &  0.46 \\ 
36356 & $-$1.05 & 0.30 & 0.26   &  0.44 & 0.45 & 0.50 & 0.28 &  0.27 & 0.43 & $-$0.02 &    0.03 &  0.51 \\ 
36929 & $-$1.03 & 0.34 & 0.45   &  0.41 & 0.73 & 0.50 & 0.30 &  0.28 & 0.38 & $-$0.04 &    0.01 &  0.46 \\ 
36942 & $-$0.98 & 0.55 & 0.04   &  0.52 & 0.37 & 0.45 & 0.24 &  0.27 & 0.37 & $-$0.16 &    0.07 &  0.45 \\ 
37215 & $-$1.11 & 0.45 & 0.25   &  0.50 & 0.62 & 0.48 & 0.29 &  0.27 & 0.47 & $-$0.07 & $-$0.02 &  0.38 \\ 
38075 & $-$1.07 & --   & 0.42   &  0.53 & 0.71 & 0.54 & 0.33 &  0.27 & 0.31 & $-$0.05 &    0.08 &  0.52 \\ 
38383 & $-$1.10 & 0.39 & 0.11   &  0.54 & 0.39 & 0.44 & 0.25 &  0.27 & 0.26 & $-$0.06 & $-$0.01 &  0.18 \\ 
38399 & $-$1.08 & 0.52 & 0.19   &  0.55 & 0.64 & 0.50 & 0.29 &  0.30 & 0.32 & $-$0.03 &    0.04 &  0.42 \\ 
38896 & $-$1.02 & 0.31 & 0.43   &  0.48 & 0.49 & 0.47 & 0.28 &  0.33 & 0.39 & $-$0.04 &    0.01 &  0.49 \\ 
42490 & $-$1.07 & 0.29 & 0.38   &  0.48 & 0.58 & 0.50 & 0.24 &  0.29 & 0.33 & $-$0.09 &    0.03 &  0.43 \\ 
42620 & $-$1.09 & 0.22 & 0.48   &  0.49 & 0.64 & 0.52 & 0.29 &  0.26 & 0.33 & $-$0.05 &    0.03 &  0.45 \\ 
43370 & $-$1.05 & --   & 0.34   &  0.40 & 0.51 & 0.51 & 0.29 &  0.28 & 0.44 & $-$0.02 &    0.02 &  0.31 \\ 
44243 & $-$1.04 & --   & 0.41   &  0.39 & 0.47 & 0.53 & 0.22 &  0.25 & 0.49 & $-$0.04 &    0.01 &  0.43 \\ 
44595 & $-$1.07 & 0.20 & 0.44   &  0.52 & 0.61 & 0.46 & 0.29 &  0.31 & 0.38 & $-$0.00 &    0.03 &  0.34 \\ 
44616 & $-$1.04 & 0.28 & 0.44   &  0.53 & 0.65 & 0.46 & 0.29 &  0.25 & 0.33 &    0.01 &    0.02 &  0.37 \\ 
45163 & $-$1.10 & --   & 0.46   &  0.49 & 0.57 & 0.54 & 0.28 &  0.21 & 0.35 & $-$0.03 & $-$0.00 &  0.43 \\ 
45895 & $-$1.04 & 0.43 & 0.10   &  0.57 & 0.54 & 0.46 & 0.26 &  0.23 & 0.32 & $-$0.07 &    0.00 &  0.48 \\ 
 5359 & $-$1.03 & 0.42 & 0.13   &  0.55 & 0.53 & 0.50 & 0.31 &  0.30 & 0.39 & $-$0.03 &    0.03 &  0.24 \\
\end{longtable}
}

\end{document}